\documentclass[aps,10pt,nofootinbib,groupedaddress,preprintnumbers,superscriptaddress]{revtex4}

\usepackage{amssymb, amsfonts, amsmath, bm}
\usepackage
{graphicx}
\usepackage
{hyperref}
\usepackage{color}

\usepackage{enumerate}

\hypersetup{%
 setpagesize=false,
 bookmarksnumbered=true,%
 bookmarksopen=true,%
 colorlinks=true,%
 linkcolor=blue,%
 citecolor=red}

\def\ben{\begin{equation}}
\def\een{\end{equation}}
\def\bena{\begin{eqnarray}}
\def\eena{\end{eqnarray}}
\def\non{\nonumber}
\def\mS{{\mathbb S}}
\def\mV{{\mathbb V}}
\def\mT{{\mathbb T}}

\begin{document}
\title{
{\large
Massive tensor field perturbations on extremal and near-extremal static black holes
}}
\author{Vitor Cardoso}
\email{vitor.cardoso@ist.utl.pt}
\affiliation{CENTRA, Departamento de F\'{\i}sica, Instituto Superior
T\'ecnico, Universidade de Lisboa,
Avenida Rovisco Pais 1, 1049 Lisboa, Portugal}

\author{Takahisa Igata}
\email{igata@rikkyo.ac.jp}
\affiliation{Department of Physics, Rikkyo University, Toshima, Tokyo 171-8501, Japan}

\author{Akihiro Ishibashi}
\email{akihiro@phys.kindai.ac.jp}
\affiliation{Department of Physics, and Research Institute for Science and Technology, 
Kindai University, Higashi-Osaka, Osaka, 577-8502, Japan} 

\author{Kodai Ueda}
\email{kodaiueda01@gmail.com}
\affiliation{Department of Physics, Kindai University, Higashi-Osaka, Osaka, 577-8502, Japan}

\date{\today}
\preprint{RUP-19-12}

\begin{abstract}
We develop a new perturbation method to study the dynamics of massive tensor fields on
extremal and near-extremal static black hole spacetimes in arbitrary dimensions. On such backgrounds, one can classify
the components of massive tensor fields into the tensor, vector, and scalar-type components. 
For the tensor-type components, which arise only in higher dimensions, the massive tensor field equation reduces to a single master
equation, whereas the vector and scalar-type components remain coupled. We consider the near-horizon expansion of both the geometry and the field variables with respect to the near-horizon scaling parameter. 
By doing so, we reduce, at each order of the expansion, the equations of motion for the vector and scalar-type components 
to a set of five mutually decoupled wave equations
with source terms consisting only of the lower-order variables. Thus, together with the tensor-type master
equation, we obtain the set of mutually decoupled equations at each order of the expansion that govern all dynamical
degrees of freedom of the massive tensor field on the extremal and near-extremal static black hole background. 
\end{abstract}

\maketitle


\section{Introduction}
\label{sec:1}

One of the most challenging problems in modern cosmophysics is to identify the dark sector of our universe. A number of appealing dark sector candidates have been proposed, including ultralight-bosons and axion-like particles in string theory-motivated models~\cite{Arvanitaki:2009fg,Acharya:2015zfk,Goodsell:2009xc,Arvanitaki:2010sy}. 
It is desirable to exploit current and future astrophysical observations, such as the newly-born gravitational wave astronomy, to hunt for the dark sector particles. In particular, as a source of gravitational waves from extremely strong gravity region, astrophysical black holes can be an excellent laboratory for understanding dark sector physics, as well as 
for testing various theories of gravity. This can be done, for example, by observing the phenomena called ``superradiant instability" or ``black hole bomb" \cite{Press:1972zz} which can occur through the interaction between a rotating black hole and some ultralight bosonic fields~\cite{Arvanitaki:2009fg}. 
It is therefore important to analyze the dynamics of various kinds of massive bosonic fields 
in black hole spacetimes~\cite{Damour:1976kh,Zouros:1979iw,Detweiler:1980uk,Dolan:2007mj,
Rosa:2009ei,Hod:2011,Pani:2012vp,Pani:2012bp,Witek:2012tr,Brito:CP:2013,Yoshino:Kodama:2015,Hod:2016,Ishibashi:Pani:Gualtieri:Cardoso:2015} (See also~\cite{Cardoso:2004nk,Herdeiro:DR2013,Hod:2013a,Degollado:Herdeiro:2014,LiZhao:2015,Hawking:Reall:2000,Cardoso:2004hs,Cardoso:2006wa,
Uchikata:2009zz,Kodama:2009rq,Cardoso:Dias:Hartnett:Lehner:Santos:2014,Green:Hollands:Ishibashi:Wald:2016,Cardoso:2018tly} for related works on the superradiant instability. Also related is the interesting possibility of hairy black holes~\cite{Herdeiro:Radu2014,Herdeiro:Radu2017,Ganchev:Santos:2017,Degollado:Herdeiro:Radu:18}.) 

\par 
A number of analyses for the dynamics of scalar, electromagnetic, and gravitational fields have been made, thanks to the separability of the corresponding Teukolsky wave equations on the Kerr black hole background. 
However, for massive vector fields (``Proca" or massive spin-$1$ particles) and massive tensor fields (i.e., massive spin-$2$ particles), the gauge-freedom is no longer available and their equations of motion in a curved background are complicated with intricately coupled dynamical variables so that even their separability is far from 
obvious. (See Ref.~\cite{Frolov:Krtous:Kubiznak:Santos:2018} for recent progress in the separability of Proca equations in the Kerr background 
and the original proposal of the new ansatz for Maxwell fields~\cite{Lunin:2017}.) 
For this reason, most studies for these massive fields in black hole spacetimes have been performed by resorting to
numerical computations~\cite{Zilhao:Witek:Cardoso:2015,East:Pretorius:2017,Dolan2018,East2018}.  
However, in order to verify the accuracy of numerical calculations as well as to understand precisely the behavior of massive fields in black hole spacetimes, it is important to develop some analytic approach for studying the massive field dynamics. 

\par 
Such a complexity for massive fields is in fact the case not only on rotating black hole backgrounds but also on {\em static} black hole 
spacetimes~\cite{Konoplya:2006,Konoplya:Zhidenko:Molina:2007,Rosa:Dolan:2012}. 
It is also worth noting that even in a static black hole, if it is charged, then superradiant scattering can occur for charged 
fields (see, e.g., a review \cite{BCP15:review:superradiance} and references therein). Therefore in order to get insights into the rotating case, as well as to explore further applications in black hole physics, it is of considerable interest in developing a useful formalism for the analytic study of massive vector and tensor field dynamics in a static black hole background. Such a formalism--besides the astrophysical context--may also be useful in the context of string theories and holographic scenarios, in which asymptotically AdS black holes and various effectively massive fields play a role.  
For massive vector fields, such a formalism has recently been developed in static, extremal and near-extremal black holes~\cite{Ueda:2018xvl}. 
The purpose of the present paper is to generalize the formalism of Ref.~\cite{Ueda:2018xvl} to the case of massive {\em tensor} fields 
on the same type of static black hole spacetimes.  
%

\par 
In this paper, following the strategy of Ref.~\cite{Ueda:2018xvl}, we first consider as our backgrounds, a generic class of warped product metrics ${\bar g}_{\mu \nu}$ that include static black holes in general $D$-dimensions and massive tensor fields $h_{\mu \nu}$ 
that obey linear wave equations on the background ${\bar g}_{\mu \nu}$. 
One may view either that both ${\bar g}_{\mu \nu}$ and $h_{\mu \nu}$ obey the same massive gravity theory with $h_{\mu \nu}$ 
being a linearized perturbation off of ${\bar g}_{\mu \nu}$, or that the background ${\bar g}_{\mu \nu}$ is given by, 
say, general relativity, while $h_{\mu \nu}$ is a test/probe field which lives in the dark sector or which acquires the mass term effectively by, 
e.g., string compactifications. 
We can classify the linear massive tensor field $h_{\mu \nu}$ on the background geometry ${\bar g}_{\mu \nu}$ into three different types: the tensor-, vector-, and scalar-types. 
For each type of the variables, by separating the ``angular'' coordinates, we reduce the equations of motion for $h_{\mu \nu}$ 
to a set of equations on a two-dimensional spacetime spanned by the time and radial coordinates. 
At this stage, the equations of motion for $h_{\mu \nu}$ are still intricately coupled. As the next step, we restrict our attention to 
extremal and near-extremal black holes. It is known that such a black hole admits the so-called {\em near-horizon geometry} obtained 
by taking a scaling limit of the neighborhood of the event horizon~\cite{Bardeen:Horowitz:1999}. 
The near-horizon geometry in general possesses enhanced isometries higher than the isometries of its original black hole~\cite{KLR07} and 
has been extensively studied in various contexts~\cite{Figueras:Kunduri:Lucietti:Rangamani:2008,Kunduri:Lucietti:2009,Hollands:Ishibashi:2010,Kunduri:Lucietti:2013}, including e.g., stability of extremal black holes~\cite{Dias:Monteiro:Reall:Santos:2010,Aretakis:2011,Durkee:Reall:11,TanahashiMurata12,LuciettiMurataReallTanahashi12,MurataReallTanahashi13,
Hollands:Ishibashi:15,Zimmerman:2017} and applications of the AdS-CFT correspondence~\cite{Gubser2008,Hartnoll:Herzog:Horowitz:2008,Guica:Hartman:Song:Strominger:2009,Porfyriadis:Strominer:2014,Porfyriadis:Shi:Strominger:2017}. 

We show that on the near-horizon geometry, the equations of motion for the massive tensor field $h_{\mu \nu}$ reduce to a set of mutually decoupled master equations. Then, by viewing the scaling parameter, say $\lambda$, as a small perturbation parameter, we expand both the field variable $h_{\mu \nu}$ and the background metric ${\bar g}_{\mu \nu}$ in power series in $\lambda$ about $\lambda=0$ as 
\begin{eqnarray}
h_{\mu \nu} &=& h^{(0)}_{\mu \nu} + \lambda h^{(1)}_{\mu \nu} + \lambda^2 h^{(2)}_{\mu \nu}+ \cdots \,, 
\\ 
{\bar g}_{\mu \nu} &=& {\bar g}^{(0)}_{\mu \nu} + \lambda {\bar g}^{(1)}_{\mu \nu} + \lambda^2 {\bar g}^{(2)}_{\mu \nu} +\cdots \,,
\end{eqnarray}  
with the leading metric ${\bar g}^{(0)}_{\mu \nu} = {\bar g}_{\mu \nu}|_{\lambda=0} $ being the near-horizon geometry. 
By doing so, one can successively analyze the dynamics even in far region away from the horizon. 
Finally, we show that at each order of $\lambda$ (and for each type of perturbations), the equations of motion 
reduce to a set of decoupled equations with source terms given by the lower-order variables.

\par 
In the next section, we describe a warped product spacetime in $D=m+n$-dimensions as our background, provide some geometric formulas used in the subsequent sections, and 
write down the equations of motion for massive tensor fields in our warped product background, thereby establishing our notation. 
Then, in Sec.~\ref{sec:3},  we decompose massive tensor fields into three different types: the tensor-, vector-, and scalar-type with respect to the $n$-dimensional base space.  By introducing harmonic tensors defined on the $n$-dimensional base space for each type of variables, we reduce the equations of motion for massive tensor fields to equations 
in $m$-dimensional spacetime. 
Sections \ref{sec:2} and \ref{sec:3} largely follow Refs.~\cite{KIS00,Kodama:Ishibashi:2003}. 
In Sec.~\ref{sec:4}, as a concrete background example we consider $4$-dimensional extremal and near-extremal Reissner--Nordstrom black hole. 
By taking the near-horizon limit and expanding both the background and the field variables with respect to the near horizon scaling parameter as explained above, we derive 
a set of decoupled master wave equations at each order of the near-horizon expansion. 
In Sec.~\ref{sec:5}, the same method applies to more generic class of static extremal and near-extremal 
black holes. Section~\ref{sec:6} is devoted to summary and discussion. We give some concrete expressions of some formulas in Appendix.

\section{Massive tensor fields in a generic warped product spacetime}
\label{sec:2}
With an eye to a wide variety of applications in fundamental physics, we shall start with a fairly generic class of warped product background spacetimes, 
provide some geometric formulas, and write down the equations of motion for massive tensor fields in our background. 

\subsection{Background geometry} 

Consider $D=(m+n)$-dimensional spacetime $(\mathcal{M}, \bar{g}_{\mu \nu})$ that is given as a warped product of 
an $m$-dimensional spacetime $(\mathcal{N}^m, g_{ab})$ and an $n$-dimensional space $(\mathcal{K}^n, \gamma_{ij})$ 
with the metric form 
\begin{align}
\label{eq:met}
\bar{g}_{\mu\nu}\:\!\mathrm{d}x^\mu\:\!\mathrm{d}x^\nu
= g_{ab}(y)\:\!\mathrm{d}y^a \:\!\mathrm{d}y^b
+r^2(y)\:\!\gamma_{ij}(z)\:\!\mathrm{d}z^i\:\!\mathrm{d}z^j \,,  
\end{align}
where $y^a$ and $z^i$ are local coordinates in $\mathcal{N}^m$ and $\mathcal{K}^n$, respectively.  
Accordingly, we hereafter use the greek indices to denote geometric quantities on $\cal M$, the latin indices in the range $a$, $b$, $c$, $\ldots$ on $\mathcal{N}^m$, and the latin indices in the range $i$, $j$, $k$, $\ldots$ on $\mathcal{K}^n$. 
We may occasionally use the latin indices to denote the corresponding coordinate components of geometric quantities 
on $\mathcal{M}$. 
We assume that $(\mathcal{K}^n, \gamma_{ij})$ be an Einstein space so that its Ricci tensor $\hat{R}_{ij}$ is given by 
\ben
 \hat{R}_{ij} = K(n-1)\gamma_{ij} \,, 
\een
where $K$ is some constant and hereafter set to be normalized as $K=0, \pm 1$.

\medskip
Let $\bar{\nabla}_\mu$, $D_a$, and $ \hat{D}_i$ be the covariant derivatives 
associated with $\bar{g}_{\mu \nu}$, $g_{ab}$, and ${\gamma}_{ij}$, respectively. Then, we find that 
the nonvanishing components of the Christoffel symbol associated with $\bar{g}_{\mu \nu}$ are given by 
\ben
 \bar{\mit \Gamma}^a{}_{bc} = {\mit \Gamma}^a{}_{bc} \,, \quad 
 \bar{\mit \Gamma}^a{}_{ij} = -r(D^a r) \gamma_{ij} \,, \quad 
 \bar{\mit \Gamma}^i{}_{aj} = \frac{D_a r}{r}\delta^i{}_j \,, \quad 
 \bar{\mit \Gamma}^i{}_{jk} = \hat {\mit \Gamma}^i{}_{jk} \,, \quad 
\label{def:Christoffel}
\een
where $\mit \Gamma^a{}_{bc}$ and $\hat{\mit \Gamma}^i{}_{jk}$ are the Christoffel symbols associated with 
$g_{ab}$ and $\gamma_{ij}$, respectively. We also find that the nonvanishing components of 
the curvature tensor $\bar{R}^\mu{}_{\nu \lambda \sigma}$ on $\cal M$ are related to 
the curvature tensors $R^a{}_{bcd}$ on $\mathcal{N}^m$ and $\hat{R}^i{}_{jkl}$ on $ \mathcal{K}^n$ as 
\ben
\bar{R}^a{}_{bcd} = R^a{}_{bcd} \,,  \quad 
\bar{R}^a{}_{ibj}= - r{D^aD_br} \gamma_{ij} \,, \quad 
\bar{R}^i{}_{jkl}= \hat{R}^i{}_{jkl}-(D^cr)(D_cr)\left(\delta^i{}_k\gamma_{jl}-\delta^i{}_l \gamma_{jk} \right) \,. 
\label{comp:curvature}
\een 
Note that when $(\mathcal{K}^n, \gamma_{ij})$ is a constant curvature space, $\hat{R}^i{}_{jkl}= K\left(\delta^i{}_k\gamma_{jl}-\delta^i{}_l \gamma_{jk} \right)$ with $K$ corresponding to its sectional curvature. 
The nonvanishing components of the Ricci curvature $\bar{R}_{\mu\nu}:=\bar{g}^{\alpha\beta}R_{\mu\alpha\nu\beta}$ are given by
\begin{align}
\bar{R}_{ab}=R_{ab}-n\:\!\frac{D_a D_b r}{r},
\ 
\bar{R}_{ai}=0,
\ 
\bar{R}_{ij}=\left[\:\!
(n-1) \frac{K-(D_c r)(D^c r)}{r^2}-\frac{D^c D_c r}{r}
\:\!\right]\bar{g}_{ij},
\end{align}
where $R_{ab}:=g^{cd}R_{acbd}$.
The scalar curvature $\bar{R}$ is 
\begin{align}
\bar{R}=R-(n+1)\:\!\frac{D^c D_c r}{r}+n(n-1)\:\!\frac{K-(D_c r)(D^c r)}{r^2},
\end{align}
where $R:=g^{ab}R_{ab}$. 

\subsection{Equations for massive tensor fields in a generic warped product spacetime}

We consider the following equations of motion for massive tensor field $h_{\mu \nu}$ with the mass-squared $\mu^2$, 
\begin{align}
\label{eq:eom}
&\bar{\Box}\:\! h_{\mu\nu}+2 \bar{R}_{\alpha\mu\beta\nu}h^{\alpha\beta}
 -\bar{R}^\alpha{}_\mu h_{\alpha \nu}
 -\bar{R}^\alpha{}_\nu h_{\alpha \mu}
-\mu^2 h_{\mu\nu}=0 \,,
\end{align} 
accompanied with the conditions: 
\begin{align}
\label{eq:consdiv}
&\bar{\nabla}_\nu h^\nu{}_{\mu}=0 \,,
\\
\label{eq:constr}
&h^\mu{}_\mu=0 \,, 
\end{align} 
where $\bar{\Box}=\bar{\nabla}^\mu \bar{\nabla}_\mu$. The massive tensor field $h_{\mu \nu}$ possesses $(D-2)(D+1)/2$ independent components. 
The action and derivation of the above equations for massive tensor fields are given in Refs.~\cite{FP1939,Hinterbichler2012} (See also Ref.~\cite{Brito:CP:2013}).  
As mentioned before, one can view the above massive tensor field as a probe field propagating on a spacetime given by the standard Einstein gravity or 
as a linear perturbation of a solution to some nonlinear massive gravity theory, such as the nonlinear massive gravity~\cite{dRGT11,dR14}.    

\medskip 
By using the formulas~(\ref{def:Christoffel}) and (\ref{comp:curvature}), the $(a,b)$-, $(a,j)$-, and $(i,j)$-components of Eq.~\eqref{eq:eom} are written, respectively, as 
\begin{align}
\label{eq:eom_ab}
&D^c D_c h_{ab}
+\frac{\hat{\Delta}}{r^2}h_{ab}
+n\:\!\frac{D^c r}{r} \left[\:\!
D_c h_{ab}-\frac{D_a r}{r}h_{bc}
-\frac{D_b r}{r} h_{ac}
\:\!\right]
-\frac{2}{r^2}\left[\:\!
\frac{D_a r}{r} \hat{D}^i h_{bi}
+\frac{D_b r}{r} \hat{D}^i h_{ai}
\:\!\right]
+2 R_{acbd} h^{cd}
\cr
&~~~~~~
-R^c{}_a h_{cb}
-R^c{}_b h_{ca}
+\frac{n}{r} \left[\:\!
(D^c D_a r) h_{cb}+(D^c D_b r)h_{ca}
\:\!\right]
-\mu^2 h_{ab}
+2\left[\:\!
\frac{(D_a r)(D_b r)}{r^2}-\frac{D_a D_b r}{r}
\:\!\right]\bar{g}^{ij} h_{ij}=0 \,, 
\end{align}
\begin{align}
\label{eq:eom_aj}
&2\:\!\frac{D^c r}{r} \hat{D}_j h_{ac}
+r D^c D_c \left(
\frac{h_{aj}}{r}
\right)
+n \:\!(D^c r)D_c \left(\frac{h_{aj}}{r}\right)
-\frac{(D_c r)(D^c r)}{r^2} h_{aj}
-\left[\:\!\mu^2-\frac{\hat{\Delta}}{r^2}\:\!\right]h_{aj}
-\frac{2}{r^2}\frac{D_a r}{r} \hat{D}^i h_{ij}
\cr
&~~~~~
-\left[\:\!
-\frac{D^c D_c r}{r}+(n-1)\frac{K-(D_c r)(D^c r)}{r^2}
\:\!\right]h_{aj}
-\left[\:\!
(n+2) \:\!\frac{(D_a r)(D^c r)}{r^2}-(n+2) \:\!\frac{D_a D^c r}{r}
+R^c{}_a
\:\!\right]h_{cj}=0 \,, 
\end{align}

\begin{align}
\label{eq:eom_ij}
&2\:\!\bar{g}_{ij} \left[\:\!
\frac{(D^a r)(D^br) }{r^2}-\frac{D^a D^b r}{r}
\:\!\right]h_{ab}
+2\:\!\frac{D^a r}{r}  (\hat{D}_i h_{aj}+\hat{D}_j h_{ai})
+r^2 D^c D_c \left(\frac{h_{ij}}{r^2}\right)
+n\:\!\frac{D^a r}{r} \:\!r^2 D_a \left(\frac{h_{ij}}{r^2}\right)
\cr
&~  
+2 r^2 \hat{R}_{ikjl}h^{kl}-2 \bar{g}_{ij} \frac{(D^cr)(D_cr)}{r^2} h^k{}_k 
+ \left[
           \frac{\hat{\Delta}}{r^2}+2\frac{D^cD_cr}{r} + 2(n-1)\frac{(D^cr)(D_cr)}{r^2}-2(n-1)\frac{K}{r^2} -\mu^2
   \right] h_{ij} = 0 \,,
\end{align}
where $\hat{\Delta} = \gamma^{ij}\hat{D}_i \hat{D}_j$. Note that the indices of $h^{kl}$ and $h^k{}_k$ in Eq.~(\ref{eq:eom_ij}) 
are raised by $\bar{g}^{ij}$. The constraints \eqref{eq:consdiv} and \eqref{eq:constr} are decomposed as
\begin{align}
\label{eq:nablaha}
&\frac{1}{r^n} D^b(r^n h_{ab})+\frac{1}{r^2} \hat{D}^i h_{ai}-\frac{D_a r}{r} \bar{g}^{ij} h_{ij}=0 \,, 
\\
\label{eq:nablahi}
&\frac{1}{r^n} D^a (r^n h_{ai}) +\frac{1}{r^2} \hat{D}^j h_{ji}=0 \,,
\\
\label{eq:haahii}
&h^a{}_a+h^{i}{}_i =0 \,.
\end{align}

\section{Decomposition of tensor fields and equations of motion} 
\label{sec:3}

In general, the components of any second rank symmetric tensor field on ${\cal M}={\cal N}^m \times {\cal K}^n$ can be decomposed 
into the three different types---called the {\em tensor}, {\em vector}, and {\em scalar}-type---according to their tensorial behavior 
on ${\cal K}^n$~\cite{KIS00,Kodama:Ishibashi:2003}. 
By introducing harmonic tensors on ${\cal K}^n$ and expanding the tensor field variables in terms of them, 
one can reduce the equations of motion for each type of the tensor fields into a set of equations on ${\cal N}^m$. 
For simplicity, throughout this paper we shall omit the index for labeling the harmonics and the summation symbol with respect to the index. 

\subsection{Tensor-type component}
We first consider the tensor-type components of $h_{\mu\nu}$ which can be expanded in terms of the harmonic tensor fields 
$\mathbb{T}_{ij}$ on $\mathcal{K}^n$ defined by 
\begin{align}
 \hat{\Delta}_{\mathrm{L}} \mathbb{T}_{ij} = \lambda_{\rm L} {\mathbb T}_{ij} \,,
 \quad
        {\hat D}^i {\mT}_{ij} = 0\,, 
\quad 
        {\mT}^i{}_i = 0 \,,
\end{align}  
where $\lambda_{\rm L}$ is the eigenvalue of the Lichnerowitz operator $\hat{\Delta}_{\rm L}$ on ${\cal K}^n$. 
Note that ${\hat \Delta}_{\rm L}$ is related to the Laplace--Beltrami operator $\hat{\Delta}$ on ${\cal K}^n$ as 
\ben
\hat{\Delta}_{\mathrm{L}} \mathbb{T}_{ij} =-\hat{\Delta}\mathbb{T}_{ij}-2\:\!\hat{R}_{ikjl}\mathbb{T}^{kl}
+2\:\!K (n-1)\mathbb{T}_{ij} \,.  
\een 
Therefore when $({\cal K}^n,\gamma_{ij})$ is a constant curvature space and $\mathbb{T}_{ij}$ is the eigentensor of $\hat \Delta $ 
with eigenvalue $k^2_{\rm T}$ (i.e., $(\hat{\Delta}+k_{\mathrm{T}}^2) \mathbb{T}_{ij}=0$), we have $\lambda_{\rm L} = k_{\rm T}^2 +2nK$ as discussed in Refs.~\cite{Gibbons:Hartnoll2002,Ishibashi:Kodama:2003}.  
Note also that the number of independent components of $ {\mT}_{ij}$ is ${(n-2)(n+1)}/{2}$ and only when $n \geqslant 3$, the tensor harmonics are nontrivial. 

\medskip 
In terms of $\mathbb{T}_{ij}$, the tensor-type components of $h_{\mu \nu}$ can be expanded as 
\begin{align}
h_{ab}=0, \quad 
h_{ai}=0, \quad
h_{ij}=2\:\!r^2 H_{\mathrm{T}} \mathbb{T}_{ij},
\end{align}
where $H_{\mathrm{T}}$ is a scalar on $(\mathcal{N}^m, g_{ab})$. 
Then we obtain a nontrivial equation for $H_{\mathrm{T}}$ from Eq.~\eqref{eq:eom_ij} 
\begin{align}
\label{eq:Teqij}
\left[\:\!
D^c D_c+n\:\!\frac{D^c r}{r} D_c+2\:\!\frac{D^c D_c r}{r}
+2(n-1) \frac{(D^c r)(D_c r)}{r^2}
-\mu^2-\frac{\lambda_{\mathrm{L}}}{r^2}
\:\!\right]H_{\mathrm{T}}=0 \,.
\end{align}
The tensor-type component admits only a single scalar field $H_{\rm T}$ on ${\cal N}^m$, and thus, together with the harmonic tensor field 
$\mathbb{T}_{ij}$, describes $(n-2)(n+1)/2$ independent components among the $(m+n-2)(m+n+1)/2$ dynamical degrees of freedom for $h_{\mu \nu}$.

\subsection{Vector-type component}
Next we consider the vector-type components of $h_{\mu \nu}$. We introduce the harmonic vector fields $\mathbb{V}_i$ on ${\cal K}^n$ 
which satisfy
\ben
 ({\hat \triangle} + {k}_{\mathrm{V}}^2 ) {\mV}_{i} = 0 \,, 
 \quad
        \hat{ D}^i {\mV}_i = 0\,, 
\een
where the eigenvalue $k_{\mathrm{V}}^2$ is given, for instance, when ${\cal K}^n$ is the unit $n$-sphere, by 
$k_{\mathrm{V}}^2= l(l+n-1) - 1\,, \; l=1,2, \dots$. The number of independent components of 
$ {\mV}_i$ is $n-1$ and only when $n \geqslant 2$, $ {\mV}_i$ is nontrivial. 
We also define the symmetric tensor $ {\mV}_{ij} $ by 
\ben
 {\mV}_{ij} = - \frac{1}{2k_{\mathrm{V}}} ( \hat{D}_i{\mV}_j + \hat{D}_j{\mV}_i) \,, 
\een
which satisfies
\ben
 \hat{\Delta}_{\rm L}{\mV}_{ij} = [k_{\mathrm{V}}^2-(n-1)K]{\mV}_{ij} \,.
\een

The vector-type components of $h_{\mu\nu}$ can be expanded in terms of $\mathbb{V}_i$ as 
\begin{align}
h_{ab}=0, \quad 
h_{ai}=r f_a \mathbb{V}_i, \quad
h_{ij}=2 \:\!r^2 H_\mathrm{T} \mathbb{V}_{ij},
\end{align}
where $(f_a, H_\mathrm{T})$ are a vector and a scalar field on $(\mathcal{N}^m, g_{ab})$. 
We find that Eqs.~\eqref{eq:eom_ab}, \eqref{eq:nablaha}, and \eqref{eq:haahii} 
are identically satisfied. 
From Eqs.~\eqref{eq:eom_aj}, \eqref{eq:eom_ij}, and \eqref{eq:nablahi} we obtain 
\begin{align}
\label{eq:Vconstr}
&\frac{1}{r^n} D^a (r^{n+1} f_a) +\frac{k_\mathrm{V}^2-(n-1) K}{k_\mathrm{V}}H_\mathrm{T}=0 \,, 
\\[2mm]
\label{eq:Veomaj}
 &\left[\:\! 
D^c D_c +n\:\!\frac{D^c r}{r} D_c
+\frac{D^c D_c r}{r} 
-\mu^2
-\frac{k_\mathrm{V}^2+(n-1)\:\!K-(n-2)\:\!(D_c r)(D^c r)}{r^2}
\:\!\right]f_a
\cr
&~~~~~~~~~~~~~~~~~~~~~~~~~~~~~~~~~
+\left[\:\!
(n+2)\:\!\frac{D_a D^c r}{r}-(n+2)\frac{(D_a r)(D^c r)}{r^2}
-R_a{}^c 
\:\!\right] f_c
-2\frac{D_a r}{r^2}\frac{k_\mathrm{V}^2-(n-1)K}{k_\mathrm{V}} H_\mathrm{T}=0 \,,
\\[2mm]
\label{eq:Veomij}
&
\left[\:\!
D^c D_c +n\:\!\frac{D^c r}{r} D_c
+2\:\!\frac{D^c D_c r}{r}
-\mu^2-\frac{k_\mathrm{V}^2+(n-1)K-2(n-1)(D_c r)(D^c r)}{r^2}
\:\!\right]H_\mathrm{T}-2\:\!\frac{k_\mathrm{V}}{r^2}(D^a r)f_a=0 \,.
\end{align}
We can eliminate $H_{\mathrm{T}}$ from Eq.~\eqref{eq:Veomaj} by use of Eq.~\eqref{eq:Vconstr}. Then we have
\begin{align}
\label{eq:arrVeqa}
&\left[\:\!
D^c D_c+n\:\!\frac{D^c r}{r}D_c+\frac{D^c D_c r}{r}-\mu^2-\frac{k_{\mathrm{V}}^2+(n-1) K-(n-2) (D^c r)(D_c r)}{r^2}
\:\!\right]f_a
\cr
&~~~~~~~~~~~~~~~~~~~~~~~~~~~~~~~~~~~~~~~~~~~~~
+\left[\:\!
(n+2) \frac{D_a D^c r}{r}+
n\frac{(D_a r)(D^c r)}{r^2}+2\frac{D_a r}{r} D^c-R_a{}^c
\:\!\right]f_c=0 \,.
\end{align} 
Since we have only a single constraint~(\ref{eq:Vconstr}), the vector-type components $(f_a, H_{\rm T})$ together with 
$\mathbb{V}_{i}$ describe $m (n-1)$ dynamical degrees of freedom for $h_{\mu \nu}$.  

\subsection{Scalar-type component}
We introduce the harmonic scalar fields on $\mathcal{K}^n$ by
\ben
 ({\hat \triangle} + {k}_{\mathrm{S}}^2 ) {\mS} = 0 \,. 
\een
We also define
\ben
 {\mathbb S}_i = - \frac{1}{k_{\mathrm{S}}}{\hat D}_i {\mathbb S} \,, \quad 
  {\mathbb S}_{ij} = \frac{1}{k_{\mathrm{S}}^2}{\hat D}_i{\hat D}_j {\mathbb S} + \frac{1}{n}\gamma_{ij}{\mathbb S} \,.
\een
The scalar-type components of $h_{\mu\nu}$ can be expanded by
\begin{align}
h_{ab}=f_{ab}\:\!\mathbb{S}, \quad
h_{aj}=r f_a \:\!\mathbb{S}_j, \quad
h_{ij}=2 r^2 \left(H_\mathrm{L} \gamma_{ij} \:\!\mathbb{S}+H_\mathrm{T} \:\!\mathbb{S}_{ij}\right),
\end{align}
where $(f_{ab}, f_a, H_\mathrm{L}, H_\mathrm{T})$ are a tensor, a vector, and two scalar fields on $ (\mathcal{N}^m, g_{ab})$. 
The transverse-traceless conditions~\eqref{eq:nablaha}, \eqref{eq:nablahi}, and \eqref{eq:haahii} reduce to
\begin{align}
\label{eq:sconsfaa}
&f^a{}_a+2\:\!n H_\mathrm{L}=0,
\\
\label{eq:scons:HL}
&\frac{1}{r^n} D^b (r^n f_{ab})+\frac{k_\mathrm{S}}{r} f_a -2n\:\!\frac{D_a r}{r} H_\mathrm{L}=0,
\\
\label{eq:sconstrfa}
&\frac{1}{r^n} D^a (r^{n+1}f_a)-2k_\mathrm{S} H_\mathrm{L}+2\:\!\frac{n-1}{n} \frac{k_\mathrm{S}^2-n\:\!K}{k_{\mathrm{S}}} H_\mathrm{T}=0. 
\end{align}
The field equations~\eqref{eq:eom_ab} and \eqref{eq:eom_aj}
lead to 
\begin{align}
\label{eq:Seqab}
&D^c D_c f_{ab}-\frac{k_\mathrm{S}^2}{r^2}f_{ab}
+n\:\!\frac{D^c r}{r} \left[\:\!
D_c f_{ab}-\frac{D_a r}{r} f_{bc}-\frac{D_b r}{r} f_{ac}
\:\!\right]
-\frac{2k_\mathrm{S}}{r^2}\left[\:\!
(D_a r) f_b+(D_b r) f_a
\:\!\right]
+2\:\!R_{acbd} f^{cd}
\cr
&~~~~~~~~~~~~
 -R_a{}^c f_{cb}
-R^c{}_b f_{ca}
+\frac{n}{r}\left[\:\!
(D^c D_a r) f_{cb}+(D^c D_b r) f_{ca}
\:\!\right]
-\mu^2 f_{ab}
+4\:\!n\left[\:\!
\frac{(D_a r)(D_b r)}{r^2}-\frac{D_a D_b r}{r} 
\:\!\right]H_\mathrm{L}=0,
\\[3mm]
\label{eq:Seqia}
&r D^c D_c f_a
+n (D^c r)(D_c f_a)
-\left[\:\!
\mu^2+\frac{k_\mathrm{S}^2-(n-2)(D_c r )(D^c r)}{r^2}-\frac{D^cD_c r}{r}
\:\!\right]r f_a
-\frac{2\:\!k_\mathrm{S}}{r} (D^c r) f_{ac}
\cr
&~~~~~~~~~~~~
-\left[\:\!
(n+2)\:\!\frac{(D_a r)(D^c r)}{r^2}
-(n+2)\:\!\frac{D_a D^c r}{r}+R_a{}^c
\:\!\right]r f_c
-4\:\!\frac{D_a r}{r}\left[\:\!
-k_\mathrm{S} H_\mathrm{L}+\frac{n-1}{n}\frac{k_\mathrm{S}^2-n K}{k_\mathrm{S}}H_\mathrm{T}
\:\!\right]
=0. 
\end{align}
We can eliminate $H_{\mathrm{L}}$ from Eq.~\eqref{eq:Seqab} by use of Eq.~\eqref{eq:sconsfaa} 
and can also eliminate $H_{\mathrm{L}}$ and $H_{\mathrm{T}}$ from Eq.~\eqref{eq:Seqia} by use of Eq.~\eqref{eq:sconstrfa}. 
Then, we obtain 
\begin{align}
\label{eq:Seqabarr}
&D^c D_c f_{ab}-\frac{k_\mathrm{S}^2}{r^2}f_{ab}
+n\:\!\frac{D^c r}{r} \left[\:\!
D_c f_{ab}-\frac{D_a r}{r} f_{bc}-\frac{D_b r}{r} f_{ac}
\:\!\right]
-\frac{2k_\mathrm{S}}{r^2}\left[\:\!
(D_a r) f_b+(D_b r) f_a
\:\!\right]
+2\:\!R_{acbd} f^{cd}
\cr
&~~~~~~~~~~~~
 -R_a{}^c f_{cb}
-R^c{}_b f_{ca}
+\frac{n}{r}\left[\:\!
(D^c D_a r) f_{cb}+(D^c D_b r) f_{ca}
\:\!\right]
-\mu^2 f_{ab}
-2\left[\:\!
\frac{(D_a r)(D_br)}{r^2}-\frac{D_a D_b r}{r}
\:\!\right]f^{c}{}_c=0,
\\
\label{eq:Seqaiarr}
&D^c D_c f_a
+\frac{n}{r} (D^c r)(D_c f_a)
+2\:\! \frac{D_a r}{r} D^c f_c
-\left[\:\!
\mu^2+\frac{k_\mathrm{S}^2-(n-2)(D_c r )(D^c r)}{r^2}-\frac{D^cD_c r}{r}
\:\!\right]f_a
-\frac{2\:\!k_\mathrm{S}}{r^2} (D^c r) f_{ac}
\cr
&~~~~~~~~~~~~~~~~~~~~~~~~~~~~~~~~~~~~~~~~~~~~~~~~~~~~~~~\:~~~~
+\left[\:\!
n\:\!\frac{(D_a r)(D^c r)}{r^2} 
+(n+2)\:\!\frac{D_a D^c r}{r}-R_a{}^c
\:\!\right] f_c
=0. 
\end{align}
Equation~\eqref{eq:eom_ij} provides two equations.
One of them corresponds to the trace part:
\begin{align}
\label{eq:Seqijt}
&D^c D_c H_\mathrm{L}+n\:\!\frac{D^c r}{r} D_c H_\mathrm{L}
-\left[\:\!
\mu^2+\frac{k_\mathrm{S}^2+2(D^c r)(D_c r)}{r^2}-2\:\!\frac{D^c D_c r}{r}
\:\!\right]H_\mathrm{L}
\cr
&~~~~~~~~~~~~~~~~~~~~~~~~~~~~~~~~~~~~~\:\!~~~~~~~~~~~~~~~~~~~~~~~~~~~~~~
+\left[\:\!
\frac{(D^a r)(D^b r)}{r^2}-\frac{D^a D^b r}{r}
\:\!\right]f_{ab}
+\frac{2\:\!k_\mathrm{S}}{n}
\frac{D^c r}{r^2}f_c 
=0,
\end{align}
and the other corresponds to the traceless part:
\begin{align}
\label{eq:Seqijtless}
D^c D_c H_\mathrm{T}+n\:\!\frac{D^c r}{r} D_c H_\mathrm{T}
-\left[\:\!
\mu^2+\frac{k_\mathrm{S}^2-2(n-1)(D_c r)(D^c r)}{r^2}-\frac{2 D^c D_c r}{r}
\:\!\right]H_\mathrm{T}-2\:\!k_{\mathrm{S}} \frac{D^c r}{r^2}f_c=0.
\end{align}

\medskip 

Since Eqs.~(\ref{eq:sconsfaa}), (\ref{eq:scons:HL}), and (\ref{eq:sconstrfa}) may be viewed as $m+2$ constraints on 
$m(m+1)/2 + m+1+1$ components of $(f_{ab}, f_a, H_L, H_T)$, the scalar-type components with a scalar field $\mathbb{S}$ 
describe $m(m+1)/2$ independent components of $h_{\mu \nu}$. Thus, the tensor-, vector-, and scalar-type 
components all together describe $(n-2)(n+1)/2+ m(n-1) + m(m+1)/2=(m+n-2)(m+n+1)/2$ dynamical degrees of freedom 
for the massive tensor field $h_{\mu \nu}$ on our $(m+n)$-dimensional spacetime, as should be so. 
All the equations obtained so far hold in a fairly generic class of background geometries given by the metric (\ref{eq:met}).   
In the following sections, we set $m=2$ so that our background metric~(\ref{eq:met}) describes static, extremal (and near-extremal) 
black hole. Then we apply the near-horizon expansion scheme in order to reduce the above equations to a set of decoupled equations for 
each tensorial type.

\section{$4$-dimensional Reissner--Nordstrom black hole}
\label{sec:4}

In this section, we consider, as a concrete example, $4$-dimensional extremal (and near-extremal) Reissner--Nordstrom 
black hole, and applying the near-horizon expansion mentioned before, we derive a set of decoupled equations for the vector and scalar-type.  
Note that there does not exist the tensor-type component in the four-dimensional case $n=2$. Note also that in the standard 
black hole perturbation theory, the vector- and scalar-type are sometime called the {\em axial} (or {\em odd})-mode 
and {\em polar} (or {\em even})-mode, respectively. 

\subsection{The background metric} 
Let us assume that $m=n=2$, $K=1$, and $\gamma_{ij}{\rm d}z^i {\rm d}z^j={\rm d} \Omega_{(2)}^2$ be the unit two-sphere metric. 
Then, the four-dimensional Reissner--Nordstrom metric is given in the (ingoing) Eddington--Finkelstein coordinates $y^a=(v,x),\: r=r_+ (1+x)$, as 
\ben
 {\rm d}s^2 = -F(x){\rm d}v^2 + 2 r_+ {\rm d} v {\rm d}r + r_+^2(1+x)^2 {\rm d} \Omega_{(2)}^2 \,, \quad 
 F(x) := \frac{x(x+\sigma)}{(1+x)^2} \,, 
 \label{def:metric:4RN}
\een 
where the constant $r_+$ denotes the radius of the event horizon located at $x=0$. The constant $\sigma$ is called 
the {\em extremality parameter} and related to the inner-horizon radius $r_-$ as 
\ben
 r_-= r_+(1- \sigma)\,, 
\een
so that when $\sigma \ll 1$, the metric (\ref{def:metric:4RN}) is referred to as {\em near-extremal} and when $\sigma=0$ as {\em extremal} 
black hole. 
Further, by taking the scaling transformation, 
\ben
  x \rightarrow \lambda x \,, \quad v \rightarrow \frac{r_+}{\lambda} v \,, \quad 
  \sigma \rightarrow \lambda \sigma \,,  
\een 
with the scaling parameter $\lambda>0$, the metric (\ref{def:metric:4RN}) becomes 
\begin{align}
&\bar{g}_{\mu\nu}\:\!\mathrm{d}x^\mu \:\!\mathrm{d}x^\nu
=r_+^2\:\!\left[\:\!
-F(x)\:\!\mathrm{d}v ^2+2\:\!\mathrm{d}v \:\!\mathrm{d} x+(1+\lambda x)^2\:\!\mathrm{d}\Omega_{(2)}^2
\:\!\right] \,, \quad 
F(x)=\frac{x(x+\sigma)}{(1+\lambda\:\!x)^2} \,.   
\label{def:metric:4RN:scaling}
\end{align}
The Ricci tensor of ${\cal N}^2$ is given by
\begin{align}
\label{eq:RicciRN}
R_a{}^b=-\frac{F''(x)}{2\:\!r_+^2} \delta_a{}^b.
\end{align}
We use units in which $r_+=1$ in what follows.  
On the background metric, \eqref{def:metric:4RN:scaling}, we examine the equations of motion for massive tensor 
perturbations given in the previous section. We shall perform the analysis in the vector- and scalar-type components, separately.

\subsection{Vector-type component}
We first consider the equations for the vector-type components.  
The explicit form of Eq.~\eqref{eq:Vconstr} is expressed as 
\begin{align}
\label{eq:Valpha}
&\left[\:\!
F\partial_x^2+2\:\!\partial_v \partial_x+F' \partial_x-
\mu^2-\frac{k_\mathrm{V}^2+1}{(1+\lambda\:\!x)^2}
\:\!\right]H_\mathrm{T}
+\frac{2\:\!\lambda}{1+\lambda\:\!x}(F\:\!\partial_x+\partial_v +F') H_\mathrm{T}
-\frac{2\:\!k_\mathrm{V} \lambda}{(1+\lambda\:\!x)^2} (F f_x+f_v)
\cr
&~~~~~~~~~~~~~~~~~~~~~~~~~~~~
~~~~~~~~~~~~~~~~~~~~~~~~~~~~~~~~~~~~~~~~~
~~~~~~~~~~~~~~~~~~~~~~~~~~~~~\:\!\:\!
+\frac{2\lambda^2 F}{(1+\lambda\:\!x)^2} H_\mathrm{T}=0. 
\end{align} 
The $a=(x, v)$ components of Eq.~\eqref{eq:arrVeqa} are respectively expressed as 
%
\begin{align}
\label{eq:Vbeta}
&F\:\!\partial_x^2 f_x+2\:\!\partial_x \partial_v f_x
+2\:\!F' \partial_x f_x
+\frac{2\:\!\lambda}{1+\lambda\:\!x}\left[\:\!
2\:\!\partial_v f_x+3\:\!F' f_x+2 F\partial_x f_x +\partial_x f_v
\:\!\right]
+F'' f_x
-\left[\:\!
\mu^2+\frac{k_\mathrm{V}^2+1}{(1+\lambda\:\!x)^2}\:\!\right]f_x
\cr
&~~~~~~~~~~~~~~~~~~~~~~~~~~~~~~~~~~~~~~~~~~~~~~~~~~~~~~~~~~~~~~~~~~~~~~~~~~~~~~~~~~~~~~~~~~~~~~~~
+\frac{2\lambda^2}{(1+\lambda\:\!x)^2} (F f_x+f_v)=0, 
\end{align}

%
\begin{align}
\label{eq:Vgamma}
F\:\!\partial_x^2 f_v+2\:\!\partial_v \partial_x f_v
+F'\partial_v f_x
-\left[\:\!
\mu^2+\frac{k_\mathrm{V}^2+1}{(1+\lambda\:\!x)^2}\:\!\right]f_v
+\frac{2\:\!\lambda}{1+\lambda\:\!x}\left[\:\!
\partial_v f_v+F\partial_x f_v+F' f_v
\:\!\right]=0.
\end{align}

We assume that the field variables $H_{\textrm{T}}$ and $f_a$ can be expanded in power series in $\lambda$ as 
\begin{align}
H_{\textrm{T}}
=\sum_{l=0}^\infty \lambda^l \cdot \Phi_{\textrm{V}1}^{(l)},
\quad
f_x
=\sum_{l=0}^\infty \lambda^l \cdot \Phi_{\textrm{V}2}^{(l)},
\quad
f_v
=\sum_{l=0}^\infty \lambda^l \cdot \Phi_{\textrm{V}3}^{(l)}. 
\end{align}
With these expressions, expanding Eqs.~\eqref{eq:Valpha}, \eqref{eq:Vbeta}, and \eqref{eq:Vgamma} in terms of $\lambda$, 
we obtain
\begin{align}
\label{eq:ValphaRN}
&\sum_{l=0}^\infty \lambda^l \cdot \sum_{m=0}^l \left[\:\!
L_{\alpha1}^{(m)} \Phi_{\mathrm{V}1}^{(l-m)}+
L_{\alpha2}^{(m)} \Phi_{\mathrm{V}2}^{(l-m)}+
L_{\alpha3}^{(m)} \Phi_{\mathrm{V}3}^{(l-m)}
\:\!\right]=0,
\\
\label{eq:VbetaRN}
&\sum_{l=0}^\infty \lambda^l\cdot \sum_{m=0}^l\left[\:\!
L_{\beta2}^{(m)} \Phi_{\mathrm{V}2}^{(l-m)}+
L_{\beta3}^{(m)} \Phi_{\mathrm{V}3}^{(l-m)}
\:\!\right]=0,
\\
\label{eq:VgammaRN}
&\sum_{l=0}^\infty \lambda^l\cdot \sum_{m=0}^l\left[\:\!
L_{\gamma2}^{(m)} \Phi_{\mathrm{V}2}^{(l-m)}+
L_{\gamma3}^{(m)} \Phi_{\mathrm{V}3}^{(l-m)}
\:\!\right]=0 \,.
\end{align}
Here, the differential operators $L_{\alpha I}^{(m)}$ ($I=1, 2, 3$) in Eq.~\eqref{eq:ValphaRN} are defined by 
\begin{align}
L_{\alpha1}^{(m)}
&:=(-1)^m\cdot\bigg[\:\!
(m+1) (x+\sigma) \:\!x^{m+1} \partial_x^2+2\:\!\delta_{m0} \partial_v \partial_x
+(m+1)(2\:\!x+\sigma) \:\!x^m \partial_x
+2 (\delta_{m0}-1) \:\!x^{m-1} \partial_v
\cr
&~~~~~~~~~~~~~~~~~~~~~~~~~~
-(m+1)(k_{\textrm{V}}^2+1)\:\!x^m
- \frac{m(m+1)(m+5)}{3}\:\!x^m
-\frac{m(m+1)(m+2)}{3} \sigma x^{m-1}-\delta_{m0}\mu^2
\:\!\bigg],
\\
L_{\alpha2}^{(m)}&:=(-1)^m\cdot \frac{m(m+1)(m+2)}{3} k_{\textrm{V}}(x+\sigma) \:\!x^m,
\\
L_{\alpha3}^{(m)}&:=(-1)^m \cdot
2\:\!k_{\textrm{V}} m \:\!x^{m-1},
\end{align}
the differential operators $L_{\beta I}^{(m)}$ ($I=2, 3$) in Eq.~\eqref{eq:VbetaRN} are defined by 
\begin{align}
L_{\beta2}^{(m)}&:=
(-1)^m \cdot \bigg[\:\!
(m+1)(x+\sigma)\:\!x^{m+1}\partial_x^2
+2\:\!\delta_{m0} \partial_x \partial_v
+4 (\delta_{m0}-1)\:\!x^{m-1} \partial_v 
+4(m+1)\:\!x^{m+1} \partial_x
+2\:\! (m+1)\sigma \:\!x^m \partial_x
\cr&~~~~~~~~~~~~~~~~
-(m+1)(k_{\textrm{V}}^2+1)\:\!x^m -\frac{2}{3}(m+3)(m^2-1)\:\!x^m
-\frac{m(m+1)(2m+1)}{3}  \sigma\:\!x^{m-1}
-\delta_{m0}\:\!\mu^2
\:\!\bigg],
\\
L_{\beta3}^{(m)}&:=(-1)^m\cdot \left[\:\!
2\:\!(\delta_{m0}-1)\:\!x^{m-1}\partial_x
+2\:\!(\delta_{m0}+m-1)\:\!x^{m-2}
\:\!\right],
\end{align}
and the differential operators $L_{\gamma I}^{(m)}$ ($I=2, 3$) in Eq.~\eqref{eq:VgammaRN} are defined by 
\begin{align}
L_{\gamma2}^{(m)}
&:=(-1)^m \cdot \bigg[\:\!
(m+1)(m+2)\:\!x^{m+1} \partial_v
+ (m+1)^2 \sigma\:\!x^m \partial_v
\:\!\bigg],
\\
L_{\gamma3}^{(m)}
&:=(-1)^m \cdot \bigg[\:\!
(m+1)(x+\sigma)\:\!x^{m+1} \:\!\partial_x^2
+2\:\!\delta_{m0} \partial_v \partial_x
+2\:\!(\delta_{m0}-1)\:\!x^{m-1} \partial_v
-m(m+1)(x+\sigma)\:\!x^m\:\! \partial_x
\cr
&~~~~~~~~~~~~~~~~~~~
-(m+1)(k_{\textrm{V}}^2+1)\:\!x^m
-\frac{2}{3}m(m+1)(m+2)\:\!x^m
-\frac{m(m+1) (2m+1)}{3}\sigma\:\!x^{m-1}
-\delta_{m0}\:\!\mu^2
\:\!\bigg]. 
\end{align}
Since all the coefficients of $\lambda^l$ must vanish, 
we obtain the equations for $\Phi_{\textrm{V} I}^{(l)}$ ($I=1, 2, 3$) in the form
\begin{align}
\left[\:\!
\begin{array}{ccc}
L_{\alpha 1}^{(0)}&0&0\\[1mm]
0&L_{\beta 2}^{(0)}& 0\\[1mm]
0&L_{\gamma 2}^{(0)}& L_{\gamma 3}^{(0)}
\end{array}
\:\!\right]\left[\:\!
\begin{array}{c}
\Phi_{\textrm{V}1}^{(l)}\\[1mm]
\Phi_{\textrm{V}2}^{(l)}\\[1mm]
\Phi_{\textrm{V}3}^{(l)}
\end{array}
\:\!\right]
=-\sum_{m=1}^l \left[\:\!
\begin{array}{ccc}
L_{\alpha 1}^{(m)}&L_{\alpha 2}^{(m)}&L_{\alpha 3}^{(m)}\\[1mm]
0&L_{\beta 2}^{(m)}& L_{\beta 3}^{(m)}\\[1mm]
0&L_{\gamma 2}^{(m)}& L_{\gamma 3}^{(m)}
\end{array}
\:\!\right]\left[\:\!
\begin{array}{c}
\Phi_{\textrm{V}1}^{(l-m)}\\[1mm]
\Phi_{\textrm{V}2}^{(l-m)}\\[1mm]
\Phi_{\textrm{V}3}^{(l-m)}
\end{array}
\:\!\right]. 
\label{eqs:4RN:vector}
\end{align}
Thus, at the leading-order, we have two mutually decoupled, homogeneous master equations 
for the two master variables $(\Phi_{\textrm{V}1}^{(0)},\Phi_{\textrm{V}2}^{(0)})$ on the near-horizon geometry ${\bar g}^{(0)}_{\mu \nu}$:
\ben
 L_{\alpha 1}^{(0)} \Phi_{\textrm{V}1}^{(0)} =0 \,, \quad 
 L_{\beta 2}^{(0)} \Phi_{\textrm{V}2}^{(0)} =0 \,.  
 \label{eqs:master:lead:vector}
\een
Once these two variables have been solved, the third variable $\Phi_{\textrm{V}3}^{(0)}$ can be determined by the two. 
Similarly, at $l(\geqslant 1)$th-order, we have two mutually decoupled inhomogeneous equations for the two master variables $(\Phi_{\textrm{V}1}^{(l)}, \Phi_{\textrm{V}2}^{(l)})$ and the third equation can be used to determine the remaining 
variable $\Phi_{\textrm{V}3}^{(l)}$. Since the source terms are given by the lower-order variables, once the leading-order 
master variables $(\Phi_{\textrm{V}2}^{(0)},\Phi_{\textrm{V}2}^{(0)})$ have been obtained, 
one can solve successively any order of the vector-type components of Eqs.~\eqref{eqs:4RN:vector} above. The two 
variables $(\Phi_{\textrm{V}1}^{(l)}, \Phi_{\textrm{V}2}^{(l)} )$ at each order together with the harmonic vector $\mathbb{V}_i$ describe two dynamical degrees of freedom, which the vector-type component should be responsible for describing.

\subsection{Scalar-type component}

We turn to consider the scalar-type components~\eqref{eq:Seqabarr}--\eqref{eq:Seqijtless}. 
First, the explicit form of the $(a,b)=(x,x)$-, $(x,v)$-, and $(v,v)$-components of Eq.~\eqref{eq:Seqabarr} are expressed, respectively, as 
\begin{align}
\label{eq:SeqxxRNpart}
&F\partial_x^2 f_{xx}
+2\:\!\partial_v \partial_x f_{xx}
+3F' \partial_x f_{xx}
+\frac{2\:\!\lambda}{1+\lambda\:\!x}(F\partial_x f_{xx}+\partial_v f_{xx})
-\frac{8\:\!\lambda^2}{(1+\lambda\:\!x)^2}f_{vx}
-\frac{4k_{\textrm{S}} \lambda}{(1+\lambda\:\!x)^2}f_{x}
\cr
&~~~~~~~~~~~~~~~~~~~~~~~~~~~~~~~~~~~~~~~~~~~~~~~~~~~~~~~~~~~~~~~
+\left[\:\!
3F''-\mu^2-\frac{k_{\textrm{S}}^2}{(1+\lambda\:\!x)^2}
-\frac{6\:\!\lambda^2 F}{(1+\lambda \:\!x)^2}
+\frac{4\:\!\lambda F'}{1+\lambda\:\!x}
\:\!\right]f_{xx}
=0 \,,
\end{align}
%
\begin{align}
&F\partial_x^2f_{xv}
+2\partial_v \partial_x f_{xv}
+F' \partial_v f_{xx}
+F' \partial_x f_{xv}
+\frac{2\lambda}{1+\lambda\:\!x}(F\partial_x f_{xv}
+\partial_v f_{xv})
+\left[\:\!
\frac{(F')^2}{2}-FF''+\frac{\lambda F F'}{1+\lambda\:\!x}
\:\!\right]f_{xx}
\cr
&~~~~~~~~~
+\left[\:\!
-\mu^2-\frac{k_{\textrm{S}}^2}{(1+\lambda\:\!x)^2}
-\frac{2\lambda^2 F}{(1+\lambda\:\!x)^2}
+\frac{4\lambda F'}{1+\lambda\:\!x}
\:\!\right]f_{xv}
-\frac{2\lambda^2}{(1+\lambda \:\!x)^2}f_{vv}
-\frac{2k_{\textrm{S}}\lambda}{(1+\lambda\:\!x)^2}f_{v}
=0 \,,
\end{align}
%
\begin{align}
&F\partial_x^2 f_{vv}
+2\:\!\partial_x \partial_v f_{vv}
-F' \partial_x f_{vv}
+2\:\!F' \partial_v f_{xv}
+\frac{2\lambda}{1+\lambda\:\!x}(F\partial_x +\partial_v)f_{vv}
+\left[\:\!
-\mu^2-\frac{k_{\textrm{S}}^2}{(1+\lambda\:\!x)^2}+F''
\:\!\right]f_{vv}
\cr
&~~~~~~~~~~~~~~~~~~~~~~~~~~~~~~~~~~~~~~~~+\left[\:\!
F'' F^2-\frac{F(F')^2}{2}
-\frac{\lambda F^2 F'}{1+\lambda\:\!x}
\:\!\right]f_{xx}
+\left[\:\!
2FF''-(F')^2
-\frac{2\:\!\lambda FF'}{1+\lambda\:\!x}
\:\!\right]f_{xv}
=0 \,. 
\end{align}

Next, the $a=x$- and $a=v$-components of Eq.~\eqref{eq:Seqaiarr} are expressed, respectively, as 
\begin{align}
\label{eq:SeqixRNmodpart}
&F\:\!\partial_x^2 f_x
+2\:\!\partial_x \partial_v f_x
+\left[\:\!
2\:\!F'+\frac{4\:\!\lambda\:\!F}{1+\lambda\:\!x}
\:\!\right]\partial_x f_x
+\frac{4\:\!\lambda}{1+\lambda\:\!x}\partial_v f_x
+\frac{2\:\!\lambda}{1+\lambda\:\!x}
\partial_x f_v
-\frac{2\:\!k_{\textrm{S}}\:\!\lambda}{(1+\lambda\:\!x)^2} (F\:\!f_{xx}+f_{vx})
\cr
&~~~~~~~~~~~~~~~~~~~~~~~~~~~~~~~~~~~~~~~~~~~+\left[\:\!
F''+\frac{6\:\!\lambda \:\!F'}{1+\lambda\:\!x}-\mu^2-\frac{k_{\textrm{S}}^2}{(1+\lambda\:\!x)^2}
+\frac{2\:\!\lambda^2 F}{(1+\lambda\:\!x)^2}
\:\!\right]f_x
+\frac{2\:\!\lambda^2}{(1+\lambda\:\!x)^2}f_v=0 \,,
\end{align}

\begin{align}
\label{eq:SeqivRNmodpart}
&F\partial_x^2 f_v
+2\:\!\partial_v \partial_x f_v
+\frac{2\lambda F}{1+\lambda\:\!x} \partial_x f_v
+F' \partial_v f_x
+\frac{2\:\!\lambda}{1+\lambda\:\!x} \partial_v f_v
-\frac{2\:\!k_{\textrm{S}}\:\!\lambda}{(1+\lambda\:\!x)^2}(F f_{xv}+f_{vv})
\cr
&~~~~~~~~~~~~~~~~~~~~~~~~~~~~~~~~~~~~~~~~~~~~~~~~~~~~~~~~~~~~~~~~~~~~~~~~~~~~~~~~~~~~~~~~
-\left[\:\!
\mu^2+\frac{k_{\textrm{S}}^2}{(1+\lambda\:\!x)^2}-\frac{2\:\!\lambda F'}{1+\lambda \:\!x}
\:\!\right] f_v=0 \,. 
\end{align}
%

Thirdly, Eqs.~\eqref{eq:Seqijt} and \eqref{eq:Seqijtless} are expressed, respectively, as 
\begin{align}
\label{SeqijtRNpart}
&(F\:\!\partial_x^2+2\:\!\partial_v \partial_x +F'\:\!\partial_x) H_{\textrm{L}}
+\frac{2\:\!\lambda}{1+\lambda\:\!x}(F\:\!\partial_x +\partial_v) H_{\textrm{L}}
-\left[\:\!
\mu^2
+\frac{k_{\textrm{S}}^2}{(1+\lambda\:\!x)^2}
+\frac{2\:\!\lambda^2 F}{(1+\lambda\:\!x)^2}
-\frac{2\:\!\lambda \:\!F'}{1+\lambda\:\!x}
\:\!\right]
H_{\textrm{L}}
\cr
&+\frac{\lambda^2F^2}{(1+\lambda\:\!x)^2}f_{xx}
+\frac{2\:\!\lambda^2 F}{(1+\lambda\:\!x)^2}f_{xv}
+\frac{\lambda^2}{(1+\lambda\:\!x)^2}f_{vv}
-\frac{\lambda \:\!F F'}{2\:\!(1+\lambda\:\!x)}f_{xx}
-\frac{\lambda\:\!F'}{1+\lambda\:\!x}f_{xv}
+\frac{k_{\textrm{S}}\:\!\lambda}{(1+\lambda\:\!x)^2}(F\:\!f_x+f_v)=0 \,, 
\end{align}
\begin{align}
\label{SeqijtlessRNpart}
&(F\:\!\partial_x^2+2\:\!\partial_v \partial_x+F'\:\!\partial_x) H_{\textrm{T}}
+\frac{2\:\!\lambda}{1+\lambda\:\!x}
(F\:\!\partial_x +\partial_v) H_{\textrm{T}}
-\left[\:\!
 \mu^2+\frac{k_{\textrm{S}}^2}{(1+\lambda\:\!x)^2}
 -\frac{2\:\!\lambda^2 F}{(1+\lambda\:\!x)^2}
 -\frac{2\:\!\lambda \:\!F'}{1+\lambda\:\!x}
\:\!\right]H_{\textrm{T}}
\cr
&~~~~~~~~~~~~~~~~~~~~~~~~~~~~~~~~~~~~~~~~~~~~~~~~~~~~~~~~~~~~~~~~~~~~~~~~~~~~~~~~~~~~~~~~~~~~~~~~~~~
-\frac{2\:\!k_{\textrm{S}} \:\!\lambda}{(1+\lambda\:\!x)^2}
(F\:\!f_x+f_v)=0 \,.
\end{align}

\medskip

Now we assume that $H_{\textrm{T}}$, $H_{\textrm{L}}$, $f_a$, $f_{ab}$ are 
expanded as the following series in $\lambda$: 
\begin{align}
&
H_{\textrm{T}}=\sum_{l=0}^\infty\lambda^l\cdot \Phi^{(l)}_{\textrm{S}1},
\quad
H_{\textrm{L}}=\sum_{l=0}^\infty\lambda^l\cdot \Phi^{(l)}_{\textrm{S}2},
\quad
f_x=\sum_{l=0}^\infty\lambda^l\cdot \Phi^{(l)}_{\textrm{S}3},
\quad
f_{xx}=\sum_{l=0}^\infty\lambda^l\cdot \Phi^{(l)}_{\textrm{S}4},
\\
&
f_v=\sum_{l=0}^\infty\lambda^l\cdot \Phi^{(l)}_{\textrm{S}5},
\quad
f_{xv}=\sum_{l=0}^\infty\lambda^l\cdot \Phi^{(l)}_{\textrm{S}6},
\quad
f_{vv}=\sum_{l=0}^\infty\lambda^l\cdot \Phi^{(l)}_{\textrm{S}7}.
\end{align}
Expanding Eqs.~\eqref{eq:SeqxxRNpart}--\eqref{SeqijtRNpart} in $\lambda$, we obtain
\begin{align}
\label{eq:SdeltaRN}
&\sum_{l=0}^\infty \lambda^l \cdot 
\sum_{m=0}^l\left[\:\!
L_{\delta 3}^{(m)} \:\!\Phi^{(l-m)}_{\textrm{S}3}
+L_{\delta 4}^{(m)} \:\!\Phi^{(l-m)}_{\textrm{S}4}
+L_{\delta 6}^{(m)} \:\!\Phi^{(l-m)}_{\textrm{S}6}
\:\!\right]=0 \,,
\\
\label{eq:SzetaRN}
&\sum_{l=0}^\infty \lambda^l \cdot 
\sum_{m=0}^l\left[\:\!
L_{\zeta 4}^{(m)} \:\!\Phi^{(l-m)}_{\textrm{S}4}
+L_{\zeta 5}^{(m)} \:\!\Phi^{(l-m)}_{\textrm{S}5}
+L_{\zeta 6}^{(m)} \:\!\Phi^{(l-m)}_{\textrm{S}6}
+L_{\zeta 7}^{(m)} \:\!\Phi^{(l-m)}_{\textrm{S}7}
\:\!\right]=0 \,,
\\
\label{eq:SetaRN}
&\sum_{l=0}^\infty \lambda^l \cdot 
\sum_{m=0}^l\left[\:\!
L_{\eta 4}^{(m)} \:\!\Phi^{(l-m)}_{\textrm{S}4}
+L_{\eta 6}^{(m)} \:\!\Phi^{(l-m)}_{\textrm{S}6}
+L_{\eta 7}^{(m)} \:\!\Phi^{(l-m)}_{\textrm{S}7}
\:\!\right]=0 \,,
\\
\label{eq:SgammaRN}
&\sum_{l=0}^\infty \lambda^l \cdot 
\sum_{m=0}^l\left[\:\!
L_{\gamma 3}^{(m)} \:\!\Phi^{(l-m)}_{\textrm{S}3}
+L_{\gamma 4}^{(m)} \:\!\Phi^{(l-m)}_{\textrm{S}4}
+L_{\gamma 5}^{(m)} \:\!\Phi^{(l-m)}_{\textrm{S}5}
+L_{\gamma 6}^{(m)} \:\!\Phi^{(l-m)}_{\textrm{S}6}
\:\!\right]=0 \,,
\\
\label{eq:SepsilonRN}
&\sum_{l=0}^\infty \lambda^l \cdot 
\sum_{m=0}^l\left[\:\!
L_{\epsilon 3}^{(m)} \:\!\Phi^{(l-m)}_{\textrm{S}3}
+L_{\epsilon 5}^{(m)} \:\!\Phi^{(l-m)}_{\textrm{S}5}
+L_{\epsilon 6}^{(m)} \:\!\Phi^{(l-m)}_{\textrm{S}6}
+L_{\epsilon 7}^{(m)} \:\!\Phi^{(l-m)}_{\textrm{S}7}
\:\!\right]=0 \,,
\\
\label{eq:SbetaRN}
&\sum_{l=0}^\infty \lambda^l \cdot 
\sum_{m=0}^l\left[\:\!
L_{\beta 2}^{(m)}\:\!\Phi_{\textrm{S} 2}^{(l-m)}
+L_{\beta 3}^{(m)}\:\!\Phi_{\textrm{S} 3}^{(l-m)}
+L_{\beta 4}^{(m)}\:\!\Phi_{\textrm{S} 4}^{(l-m)}
+L_{\beta 5}^{(m)}\:\!\Phi_{\textrm{S} 5}^{(l-m)}
+L_{\beta 6}^{(m)}\:\!\Phi_{\textrm{S} 6}^{(l-m)}
+L_{\beta 7}^{(m)}\:\!\Phi_{\textrm{S} 7}^{(l-m)}
\:\!\right]=0 \,,
\\
\label{eq:SalphaRN}
&\sum_{l=0}^\infty \lambda^l \cdot 
\sum_{m=0}^l\left[\:\!
L_{\alpha 1}^{(m)} \:\!\Phi^{(l-m)}_{\textrm{S}1}
+L_{\alpha 3}^{(m)} \:\!\Phi^{(l-m)}_{\textrm{S}3}
+L_{\alpha 5}^{(m)} \:\!\Phi^{(l-m)}_{\textrm{S}5}
\:\!\right]=0 \,.
\end{align}
Here the differential operators $L_{\alpha I}^{(m)}$ ($I=1, 3, 5$) in Eq.~\eqref{eq:SalphaRN} are defined by
\begin{align}
L_{\alpha 1}^{(m)}
&:=(-1)^m \cdot\bigg[\:\!
(m+1)(x+\sigma)\:\!x^{m+1} \partial_x^2
+2\:\!\delta_{m0}\:\!\partial_v \partial_x
+2\:\!(m+1)\:\!x^{m+1}\partial_x
+(m+1)\:\!\sigma x^{m} \partial_x
-\delta_{m0}\:\!\mu^2\cr
&~~~~~~~~~~~~~~~~~
+2\:\!(\delta_{m0}-1) x^{m-1} \partial_v
-(m+1) k_{\textrm{S}}^2x^m
-\frac{m(m+1)(m+5)}{3} x^m
-\dfrac{m(m+1)(m+2)}{3}\sigma x^{m-1}
\:\!\bigg] \,,~~~~~~~
\\
L_{\alpha 3}^{(m)}
&:=(-1)^m \cdot
\frac{m(m+1)(m+2)}{3}k_{\textrm{S}}\:\! (x+\sigma) x^m \,,
\\
L_{\alpha 5}^{(m)}
&:=(-1)^m \cdot
2\:\!m\:\!k_{\textrm{S}} \:\!x^{m-1} \,.  
\end{align}
The differential operators $L_{\beta I}^{(m)}$ ($I=2, \ldots, 7$) in Eq.~\eqref{eq:SbetaRN} 
are defined by
\begin{align}
L_{\beta 2}^{(m)}
&:=(-1)^m\cdot \bigg[\:\!
(m+1)(x+\sigma)\:\!x^{m+1}\partial_x^2
+2\:\!\delta_{m0} \partial_v \partial_x
+2\:\!(m+1)\:\!x^{m+1} \partial_x
+(m+1)\:\!\sigma x^m \partial_x
\cr
&~~~~~~~~~~~~~~~~~
+2\:\!(\delta_{m0}-1) x^{m-1} \partial_v
-\delta_{m0} \:\!\mu^2 
-(m+1)\:\!k_{\textrm{S}}^2 x^m
-m(m+1)^2 x^m
-m^2(m+1)\:\!\sigma x^{m-1} 
\:\!\bigg],
~~~~~~~
\\
L_{\beta 3}^{(m)}
&:=(-1)^m\cdot \left[\:\!
-\frac{m(m+1)(m+2)}{6}\:\! k_{\textrm{S}}\:\!(x+\sigma)\:\!x^m
\:\!\right],
\\
L_{\beta 4}^{(m)}
&:=(-1)^m\cdot 
\frac{m(m+1)(m+2)(m+3)}{2\cdot 5!}\:\!\left[\:\!
2\:\!(2m+3)x+(4m+1)\:\!\sigma
\:\!\right](x+\sigma)\:\!x^m,
\\
L_{\beta 5}^{(m)}
&:=(-1)^m\cdot 
\left(
-m \:\!k_{\textrm{S}}\:\!x^{m-1}
\right),
\\
L_{\beta 6}^{(m)}
&:=(-1)^m\cdot 
\frac{m(m+1)}{6}\left[\:\!
2\:\!(2m+1)\:\!x
+(4m-1)\:\!\sigma
\:\!\right]x^{m-1},
\\
L_{\beta 7}^{(m)}
&:=(-1)^m\cdot 
(\delta_{m0}+m-1)\:\!x^{m-2} \,. 
\end{align}
The differential operators $L_{\gamma I}^{(m)}$ ($I=3, 4, 5, 6$) in Eq.~\eqref{eq:SgammaRN} 
are defined by
\begin{align}
L_{\gamma 3}^{(m)}
&:=(-1)^m\cdot \bigg[\:\!
(m+1)(x+\sigma) x^{m+1} \partial_x^2
+2\:\!\delta_{m0} \partial_x \partial_v
+4 (\delta_{m0}-1)x^{m-1} \partial_v
+4 (m+1) x^{m+1} \partial_x
+2(m+1)\:\!\sigma x^{m} \partial_x
\cr
&~~~~~~~~~~~~~~~~~
-\frac{2}{3} (m^2-1)(m+3) x^{m}
-\frac{m(m+1)(2m+1)}{3}\sigma x^{m-1}
-(m+1) \:\!k_{\textrm{S}}^2 x^{m}
-\delta_{m0}\:\!\mu^2
\:\!\bigg] \,,~~~~~~~~~~~~~~~~~~~~~~~~~~~~~~~~~~
\\
L_{\gamma 4}^{(m)}
&:=(-1)^m\cdot
\frac{m(m+1)(m+2)}{3}\:\!k_{\textrm{S}} (x+\sigma)\:\! x^m,
\\
L_{\gamma 5}^{(m)}
&:=(-1)^m\cdot \left[\:\!
2\:\!(\delta_{m0}-1) x^{m-1} \partial_x 
+2\:\!(\delta_{m0}+m-1)\:\!x^{m-2}
\:\!\right] \,,
\\
L_{\gamma 6}^{(m)}
&:=(-1)^m\cdot 2\:\!m k_{\textrm{S}} \:\!x^{m-1} \,. 
\end{align} 
The differential operators $L_{\delta I}^{(m)}$ ($I=3, 4, 6$) in Eq.~\eqref{eq:SdeltaRN} are defined by
\begin{align}
L_{\delta 3}^{(m)}
&:=(-1)^m\cdot
4\:\!mk_{\textrm{S}} \:\!x^{m-1} \,,
\\
L_{\delta 4}^{(m)}
&:=(-1)^m\cdot
\bigg[\:\!
(m+1)(x+\sigma)x^{m+1} \partial_x^2
+2\:\!\delta_{m0}\partial_v\partial_x
+2\:\!(m+1)(m+3)x^{m+1}\partial_x
+(m+1)(2m+3)\sigma x^m \partial_x
\cr
&~~~~~~~~~~~
+2\:\!(\delta_{m0}-1)x^{m-1}\partial_v
-(m+1)k_{\textrm{S}}^2 x^m
+\frac{2}{3}(m+1)(m^2+11m+9) \:\!x^m
\cr
&~~~~~~~~~~~
+\frac{2}{3}m(m+1)(m+5)\:\!\sigma x^{m-1}
-\delta_{m0}\:\!\mu^2
\:\!\bigg] \,,~~~~~~~~~~~~~~~~~~~~~~~~~~~
\\
L_{\delta 6}^{(m)}
&:=(-1)^m\cdot
8\:\!(1-m-\delta_{m0}) \:\!x^{m-2} \,. 
\end{align}
The differential operators $L_{\epsilon I}^{(m)}$ ($I=3, 5, 6, 7$) in Eq.~\eqref{eq:SepsilonRN}
 are defined by
\begin{align}
L_{\epsilon 3}^{(m)}
&:=(-1)^m\cdot \left[\:\!
(m+1)(m+2) x^{m+1}
+(m+1)^2 \sigma x^m
\:\!\right]\partial_v \,,
\\
L_{\epsilon 5}^{(m)}
&:=(-1)^m\cdot \bigg[\:\!
(m+1) (x+\sigma) x^{m+1} \partial_x^2
+2\:\!\delta_{m0} \partial_v \partial_x
-m(m+1) (x+\sigma) x^m \partial_x
+2\:\!(\delta_{m0}-1) x^{m-1} \partial_v
-\delta_{m0}\:\!\mu^2
\cr
&~~~~~~~~~~~~~~~~~
-(m+1)k_{\textrm{S}}^2 x^m
-\frac{2}{3}m(m+1)(m+2) x^m
-\frac{m(m+1)(2m+1)}{3}\sigma x^{m-1}
\:\!\bigg] \,,
\\
L_{\epsilon 6}^{(m)}
&:=(-1)^m\cdot 
\frac{m(m+1)(m+2)}{3} k_{\textrm{S}} (x+\sigma)\:\! x^m \,,
\\
L_{\epsilon 7}^{(m)}
&:=(-1)^m\cdot 2mk_{\textrm{S}}\:\!x^{m-1} \,.
\end{align} 
The differential operators $L_{\zeta I}^{(m)}$ ($I=4, 5, 6, 7$) in Eq.~\eqref{eq:SzetaRN}
 are defined by
\begin{align}
L_{\zeta 4}^{(m)}
&:=(-1)^m\cdot \bigg[\:\!
(m+1)(m+2)x^{m+1} \partial_v
+(m+1)^2 \sigma x^m \partial_v
\cr
&~~~~~~~~~~~~
-\frac{(m+1)(m+2)(m+3)}{120}
\left[\:\!
6\:\!m(m+4)\:\!x^2
+3\:\!m(4m+11)\:\!\sigma x
+(6\:\!m^2+9\:\!m-10) \sigma^2
\:\!\right]x^m
\:\!\bigg] \,,
\\
L_{\zeta 5}^{(m)}
&:=(-1)^m\cdot 
2\:\!mk_{\textrm{S}}x^{m-1} \,,
\\
L_{\zeta 6}^{(m)}
&:=(-1)^m\cdot \bigg[\:\!
(m+1)(x+\sigma)x^{m+1} \partial_x^2
+2\:\!\delta_{m0} \partial_v \partial_x
+2\:\!(\delta_{m0}-1) x^{m-1} \partial_v
+2\:\!(m+1) x^{m+1}\partial_x
 \cr
&~~~~~~~~~~~~
+(m+1)\:\!\sigma x^m \partial_x
-(m+1)k_{\textrm{S}}^2\:\! x^m
-\frac{m(m+1)(5\:\!m+7)}{3}x^m
\cr
&~~~~~~~~~~~~
-\frac{m(m+1)(5\:\!m+1)}{3}\sigma x^{m-1}
-\delta_{m0} \:\!\mu^2\:\!\bigg] \,,
\\
L_{\zeta 7}^{(m)}
&:=(-1)^m\cdot 2 (1-m-\delta_{m0})\:\!x^{m-2} \,,
\end{align}
and the differential operators $L_{\eta I}^{(m)}$ ($I=4, 6, 7$) in Eq.~\eqref{eq:SetaRN} 
are defined by
\begin{align}
L_{\eta 4}^{(m)}
&:=(-1)^m\cdot \frac{(m+1)(m+2)(m+3)(m+4)(m+5)}{7!}
\cr
&~~~~~~~~~~~~~~~~~~~~~~~~~~~~~~~~~~~~~~~~~~\bigg[\:\!
2\:\!m(m+6)x^2+m(4m+17)\:\!\sigma x
+(2m^2+5m-21)\:\!\sigma^2
\:\!\bigg](x+\sigma)\:\!x^{m+1} \,,
\\
L_{\eta 6}^{(m)}
&:=(-1)^m\cdot \bigg[\:\!
2\:\!(m+1)(m+2) x^{m+1}\partial_v
+2\:\!(m+1)^2\sigma x^m\partial_v
\cr
&~~~~~~~~~~~~~+\frac{(m+1)(m+2)(m+3)}{60}
\left[\:\!
6\:\!m(m+4)x^2
+3\:\!m(4m+11)\sigma x
+(6\:\!m^2+9\:\!m-10)\:\!\sigma^2
\:\!\right]x^m
\:\!\bigg] \,,
\\
L_{\eta 7}^{(m)}
&:=(-1)^m\cdot \bigg[\:\!
(m+1)(x+\sigma) x^{m+1} \partial_x^2
+2\:\!\delta_{m0} \partial_x \partial_v
-2\:\!(m+1)^2 x^{m+1} \partial_x
-(m+1)(2m+1)\:\!\sigma x^m \partial_x
\cr
&~~~~~~~~~~~~~+2\:\!(\delta_{m0}-1)x^{m-1} \partial_v
-(m+1)k_{\textrm{S}}^2 x^m
+(m+1)^2\left[\:\!
(m+2) x^m+m\:\!\sigma x^{m-1}
\:\!\right]-\delta_{m0}\:\!\mu^2
\:\!\bigg] \,.
\end{align}
Since all the coefficients of $\lambda^l$ must vanish, we obtain the equations for $\Phi^{(l)}_{\textrm{S}I}$ ($I=1,2,\ldots, 7$) in the form
\begin{align}
\left[\:\!
\begin{array}{ccccccc}
L_{\alpha 1}^{(0)}&
0&
0&
0&
0&
0&
0\\[1mm]
0&
L_{\beta 2}^{(0)}&
0&
0&
0&
0&
0\\[1mm]
0&
0&
L_{\gamma 3}^{(0)}&
0&
0&
0&
0\\[1mm]
0&
0&
0&
L_{\delta 4}^{(0)}&
0&
0&
0\\[1mm]
0&
0&
L_{\epsilon 3}^{(0)}&
0&
L_{\epsilon 5}^{(0)}&
0&
0\\[1mm]
0&
0&
0&
L_{\zeta 4}^{(0)}&
0&
L_{\zeta 6}^{(0)}&
0\\[1mm]
0&
0&
0&
L_{\eta 4}^{(0)}&
0&
L_{\eta 6}^{(0)}&
L_{\eta 7}^{(0)}\\
\end{array}
\:\!\right]\left[\:\!
\begin{array}{c}
\Phi^{(l)}_{\mathrm{S}1}\\[1mm]
\Phi^{(l)}_{\mathrm{S}2}\\[1mm]
\Phi^{(l)}_{\mathrm{S}3}\\[1mm]
\Phi^{(l)}_{\mathrm{S}4}\\[1mm]
\Phi^{(l)}_{\mathrm{S}5}\\[1mm]
\Phi^{(l)}_{\mathrm{S}6}\\[1mm]
\Phi^{(l)}_{\mathrm{S}7}\\[1mm]
\end{array}
\:\!\right]
=-\sum_{m=1}^l
\left[\:\!
\begin{array}{ccccccc}
L_{\alpha 1}^{(m)}&
0&
L_{\alpha 3}^{(m)}&
0&
L_{\alpha 5}^{(m)}&
0&
0\\[1mm]
0&
L_{\beta 2}^{(m)}&
L_{\beta 3}^{(m)}&
L_{\beta 4}^{(m)}&
L_{\beta 5}^{(m)}&
L_{\beta 6}^{(m)}&
L_{\beta 7}^{(m)}
\\[1mm]
0&
0&
L_{\gamma 3}^{(m)}&
L_{\gamma 4}^{(m)}&
L_{\gamma 5}^{(m)}&
L_{\gamma 6}^{(m)}&
0\\[1mm]
0&
0&
L_{\delta 3}^{(m)}&
L_{\delta 4}^{(m)}&
0&
L_{\delta 6}^{(m)}&
0\\[1mm]
0&
0&
L_{\epsilon 3}^{(m)}&
0&
L_{\epsilon 5}^{(m)}&
L_{\epsilon 6}^{(m)}&
L_{\epsilon 7}^{(m)}
\\[1mm]
0&
0&
0&
L_{\zeta 4}^{(m)}&
L_{\zeta 5}^{(m)}&
L_{\zeta 6}^{(m)}&
L_{\zeta 7}^{(m)}
\\[1mm]
0&
0&
0&
L_{\eta 4}^{(m)}&
0&
L_{\eta 6}^{(m)}&
L_{\eta 7}^{(m)}\\[1mm]
\end{array}
\:\!\right]\left[\:\!
\begin{array}{c}
\Phi^{(l-m)}_{\mathrm{S}1}\\[1mm]
\Phi^{(l-m)}_{\mathrm{S}2}\\[1mm]
\Phi^{(l-m)}_{\mathrm{S}3}\\[1mm]
\Phi^{(l-m)}_{\mathrm{S}4}\\[1mm]
\Phi^{(l-m)}_{\mathrm{S}5}\\[1mm]
\Phi^{(l-m)}_{\mathrm{S}6}\\[1mm]
\Phi^{(l-m)}_{\mathrm{S}7}\\[1mm]
\end{array}
\:\!\right] \,. 
\label{eq:gnrl:vector}
\end{align} 
At the leading-order, we have four mutually decoupled, homogeneous equations 
for $(\Phi^{(0)}_{\textrm{S}1}, \: \Phi_{\textrm{S}2}^{(0)}, \: \Phi_{\textrm{S}3}^{(0)}, \: \Phi_{\textrm{S}4}^{(0)})$.   
Furthermore, from the transverse-traceless conditions~\eqref{eq:sconsfaa}, \eqref{eq:scons:HL}, and \eqref{eq:sconstrfa}, 
we have 
\bena
&& 2\Phi_{{\rm S}6}+ F\Phi_{{\rm S}4} + 4 \Phi_{{\rm S}2} =0 \,, 
\label{condi:traceless}
\\
&& \left( \partial_x + 2\frac{r'}{r} \right) \Phi_{{\rm S}7}
 + \left( \partial_v + F\partial_x + F'+ 2\frac{r'}{r} F \right)\Phi_{{\rm S}6}+ \frac{k_{\textrm{S}}}{r}\Phi_{{\rm S}5} =0 \,,
\label{condi:trans:1v} 
\\
&& \left( \partial_x + 2\frac{r'}{r} \right) \Phi_{{\rm S}6}
 + \left( \partial_v + F\partial_x + \frac{3}{2}F'+ 2\frac{r'}{r} F \right)\Phi_{{\rm S}4} + \frac{k_{\textrm{S}}}{r}\Phi_{{\rm S}3} - 4\frac{r'}{r}\Phi_{{\rm S}2} =0 \,,
\label{condi:trans:1x}
\\
&& \left( \partial_x + 3\frac{r'}{r} \right) \Phi_{{\rm S}5}
 + \left( \partial_v + F\partial_x + F'+ 3\frac{r'}{r} F \right)\Phi_{\textrm{S}3}-2 \frac{k_{\textrm{S}}}{r}\Phi_{\textrm{S}2} 
  + \frac{k_{\textrm{S}}^2-2K}{k_{\textrm{S}}}\frac{\Phi_{\textrm{S}1}}{r} =0 \,. 
\label{condi:trans:2} 
\eena 
Equations~\eqref{condi:traceless} and \eqref{condi:trans:1x} imply that $\Phi_{{\rm S}4}, \: \Phi_{{\rm S}6}$ are determined 
by $(\Phi_{{\rm S}2}, \: \Phi_{{\rm S}3})$, and Eq.~\eqref{condi:trans:2} implies that 
$\Phi_{{\rm S}5}$ can be determined by $(\Phi_{{\rm S}1}, \: \Phi_{{\rm S}2}, \: \Phi_{{\rm S}3})$. 
Then, $\Phi_{{\rm S}7}$ can also be determined via (\ref{condi:trans:1v}). 
Therefore, we can view the three variables $(\Phi^{(0)}_{{\rm S}1}, \: \Phi^{(0)}_{{\rm S}2}, \: \Phi^{(0)}_{{\rm S}3})$ as 
the leading-order master variables, which are governed by the homogeneous master wave equations on 
the near-horizon geometry: 
\ben
 L_{\alpha 1}^{(0)} \Phi_{\textrm{S}1}^{(0)} =0 \,, \quad 
 L_{\beta 2}^{(0)} \Phi_{\textrm{S}2}^{(0)} =0 \,, \quad   
 L_{\gamma 3 }^{(0)} \Phi_{\textrm{S}3}^{(0)} =0 \,.  
 \label{eqs:master:lead:scalar}
\een
Once these three variables have been solved, the remaining variables $(\Phi_{{\rm S}4}^{(0)}, \Phi_{{\rm S}5}^{(0)}, \Phi_{{\rm S}6}^{(0)}, \Phi_{{\rm S}7}^{(0)})$ can be determined by the three 
as explained above. 
Similarly, at $l(\geqslant 1)$th-order, we have three mutually decoupled inhomogeneous equations for the three master variables 
$(\Phi^{(l)}_{{\rm S}1}, \: \Phi^{(l)}_{{\rm S}2}, \: \Phi^{(l)}_{{\rm S}3})$, and the remaining equations can be used to determine 
the variables $\Phi_{\textrm{S}J}^{(l)}, \: J=4,5,6,7$. Since the source terms are given by the lower-order variables, once the leading-order 
master variables $\Phi_{\textrm{S}I}^{(0)},\: I=1,2,3$ have been obtained, 
one can solve successively any order of the scalar-type components. 
The three variables $\Phi_{\textrm{S}I}^{(l)}, \: I=1,2,3$ at each order together with the harmonic scalar $\mathbb{S}$ 
describe three dynamical degrees of freedom, which the scalar-type components should be responsible for describing. 


\subsection{General solutions to the leading-order master equations}

We provide the general solutions to the leading-order master equations in the extremal and near-extremal Reissner--Nordstrom backgrounds. 
Assuming the time dependency $\Phi \propto e^{-i\omega v}$, the leading-order five equations--Eqs.~\eqref{eqs:master:lead:vector} for the vector-type 
and Eqs.~(\ref{eqs:master:lead:scalar}) for the scalar-type--reduce to the following two ordinary differential equations: 
\begin{eqnarray}
&& \left[x(x+\sigma) \frac{{\rm d}^2}{{\rm d}x^2} + (2x+\sigma -2i\omega) \frac{{\rm d}}{{\rm d}x}-(A+\mu^2) \right] \Phi_{\Lambda}^{(0)} =0 \,, 
\label{eq:A}
\\
&& \left[x(x+\sigma) \frac{{\rm d}^2}{{\rm d}x^2} + 2 (2x+\sigma -2i\omega) \frac{{\rm d}}{{\rm d}x}-(B+\mu^2-2) \right] \Phi_{\Lambda}^{(0)} =0 \,, 
\label{eq:B}
\end{eqnarray}
where $\Lambda$ collectively denotes ${{\rm V}1}, {{\rm V}2}, {{\rm S}1}, {{\rm S}2}, {{\rm S}3}$. More precisely, $\Phi^{(0)}_{{\rm V}1}$, $\Phi^{(0)}_{{\rm S}1}$, and $\Phi^{(0)}_{{\rm S}2}$ obey the first Eq.~(\ref{eq:A}) with $A=k_\textrm{V}^2+1, \: k_\textrm{S}^2$ and $k_\textrm{S}^2$, respectively, while $\Phi^{(0)}_{{\rm V}2}$, $\Phi^{(0)}_{{\rm S}3}$ the second Eq.~(\ref{eq:B}) with $B=k_{\rm V}^2+1, \: k_{\rm S}^2$ respectively. 
Note that in accord with the time coordinate scaling $v \rightarrow v/\lambda$, the frequency $\omega$ is scale transformed as $\omega \rightarrow
\lambda \omega$. Equations~(\ref{eq:A}) and (\ref{eq:B}) are precisely the same as Eqs.~(79) and (80) of Ref.~\cite{Ueda:2018xvl}, respectively.  
We can immediately solve these equations and obtain the general solutions to Eq.~(\ref{eq:A}) as, 
\begin{eqnarray}
\mbox{for $\sigma \neq 0$}, \quad 
\Phi_\Lambda^{(0)}
 &=& C_1\cdot {}_2F_1\left( - \nu + \frac{1}{2}, \nu+\frac{1}{2}, 1+ 2i \frac{\omega}{\sigma}; 1 + \frac{x}{\sigma} \right) 
 \non \\
  &{}&  
     + C_2 \cdot (x+ \sigma)^{-2i \omega/\sigma} 
             {}_2F_1\left( - \nu + \frac{1}{2}- 2i \frac{\omega}{\sigma}, \nu+\frac{1}{2}- 2i \frac{\omega}{\sigma}, 1- 2i \frac{\omega}{\sigma}; 1 + \frac{x}{\sigma}  \right) \,, 
\\              
\mbox{for $\sigma =0$}, \quad 
\Phi_\Lambda^{(0)}
 &=&  \frac{1}{\sqrt{x}}e^{-i\omega/x}
   \left[
           C_1\cdot I_\nu \left( i \omega/x \right) +   C_2\cdot K_\nu \left( i \omega/x \right) 
   \right] \,,                      
\end{eqnarray}
where ${}_2F_1$, $I_\nu$, and $K_\nu$ denotes, respectively, the hypergeometric function, and the modified Bessel functions with $ \nu := \sqrt{A+ \mu^2 +1/4}$, and where $C_1,C_2$ are arbitrary constants. They are the same as Eqs.~(81) and (84) of Ref.~\cite{Ueda:2018xvl}, respectively. 
For Eq.~(\ref{eq:B}), we obtain 
\begin{eqnarray}
\mbox{for $\sigma \neq 0$}, \quad 
\Phi_\Lambda^{(0)}
 &=& C_1\cdot {}_2F_1\left( - \nu + \frac{3}{2}, \nu+\frac{3}{2}, 2 + 2i \frac{\omega}{\sigma}; 1 + \frac{x}{\sigma} \right) 
 \non \\
  &{}&  
     + C_2 \cdot (x+ \sigma)^{-1-2i \omega/\sigma} 
             {}_2F_1\left( - \nu + \frac{1}{2}- 2i \frac{\omega}{\sigma}, \nu+\frac{1}{2}- 2i \frac{\omega}{\sigma}, -2i \frac{\omega}{\sigma}; 1 + \frac{x}{\sigma}  \right) \,,              
\\
\mbox{for $\sigma =0$}, \quad 
\Phi_\Lambda^{(0)}
 &=&   
           C_1\cdot {x}^{-5/2}e^{-i\omega/x}\cdot
                          \left\{ 
                                   \omega I_{\nu +1} \left(-i \omega/x \right) 
                                   + i \left[(\nu + 1/2) x - i\omega  \right]  I_{\nu} \left(- i \omega/x \right)
                         \right\} 
\non \\
  &{+}&  C_2\cdot {x}^{-5/2}e^{-i\omega/x}\cdot
                         \left\{ 
                                  - \omega K_{\nu +1} \left(-i \omega/x \right) 
                                   + i \left[(\nu + 1/2) x - i\omega  \right]  K_{\nu} \left(-i \omega/x \right)
                         \right\} \,, 
\end{eqnarray}
where $\nu := \sqrt{B+ \mu^2 +1/4}$. See also Eqs.~(82) and (85) of Ref.~\cite{Ueda:2018xvl}, respectively. 

\par
The higher $l (\geqslant 1)$th-order equations for $\Phi^{(l)}_{\Lambda}, \: \Lambda ={{\rm V}1}, {{\rm V}2}, {{\rm S}1}, {{\rm S}2}, {{\rm S}3}$ all 
take schematically the following form
\begin{eqnarray}
   L^{(0)} \Phi^{(l)}_\Lambda = S_\Lambda^{(l)}\,, 
\end{eqnarray}  
where $L^{(0)}$ denotes the second-order differential operators appeared in the left-hand sides of Eqs.~(\ref{eq:A}) and (\ref{eq:B}), and 
where the source term $S_\Lambda^{(l)}$ in the right-hand side consists only of the lower-order variables $\Phi^{(l-m)}_\Lambda, \: (1\leqslant m \leqslant l)$. 
Now that we have the general solutions $\Phi_\Lambda^{(0)}$ to the leading-order master equations, once the boundary conditions of interest are 
determined, one can construct the Green functions, formally expressed as $G^{(0)}=L^{(0)}{}^{-1}$, 
by the standard argument and then obtain higher-order solutions at any order as $\Phi_\Lambda^{(l)}= G^{(0)}S_\Lambda^{(l)}$.

\section{General extremal and near-extremal black holes}
\label{sec:5}
In this section, we generalize our previous analyses to the case of $D=2+n$-dimensional general black holes with two horizons at $r=r_\pm$. 
The general form of the metric in Eddington--Finkelstein-type coordinates is given by
\begin{align}
\label{eq:metgen}
&\frac{1}{r_+^2}\bar{g}_{\mu\nu}\:\!\mathrm{d}x^\mu\:\!\mathrm{d}x^\nu=-F(\lambda\:\!x)\:\!\mathrm{d}v^2+2\:\!\mathrm{d}v \:\!\mathrm{d}x
+ \frac{1}{r_+^2}{r^2}(\lambda\:\!x) \gamma_{ij}\:\!\mathrm{d}z^i\:\!\mathrm{d}z^j \,, 
\\
&F(\lambda\:\!x)=x (x+\sigma)\:\!g(\lambda\:\!x) \,, 
\end{align}
where $x=0$ corresponds to the event (outer) horizon, $\lambda$ denotes the scaling parameter, $\sigma:=(r_+-r_-)/r_+$ the extremality parameter as before, 
and where $g(\lambda x)>0$ is a generic, regular and positive (nonvanishing) function. As in the Reissner--Nordstrom case, 
by taking the limit $\lambda \rightarrow 0$, we obtain the near-horizon geometry for the extremal ($\sigma=0$) and near-extremal ($ \sigma \neq 0$) 
black holes. Hereafter we normalize $r_+=1$.    

\par
We expand the background metric functions, $F(\lambda x)$ and $r (\lambda x)$ in terms of $\lambda$ as 
\begin{eqnarray}
 g(\lambda x) = \sum_{l=0}^\infty \lambda^l \cdot g^{(l)} \,, \quad r(\lambda x) = \sum_{l=0}^\infty \lambda^l \cdot r^{(l)} \,, 
\end{eqnarray}
where $g^{(l)}, r^{(l)}$ are assumed to be smooth functions of $x$ with the leading-order $g^{(0)}, r^{(0)}$ being some constants.  
For later convenience, we also introduce the following quantity $\Delta^{(l)}_{(abc;def)}$ defined by  
\begin{eqnarray}
 r^a (r')^b (r'')^c g^d (g')^e (g'')^f  = \sum_{l=0}^\infty \lambda^l \cdot \Delta^{(l)}_{(abc;def)} \,, 
\label{def:Delta:abcdef}
\end{eqnarray}
where the prime ${}'$ denotes the derivative by $x$ and $a,b,c,d,e,f$ each takes one of the values $0,1, 2$. 
The concrete expressions of $\Delta^{(l)}_{(abc;def)}$ are given in Appendix. 

\subsection{Tensor-type component}
Let us first consider the tensor-type component of the massive tensor field perturbations. This is the case when $D=2+n \geqslant 5$ since 
the harmonic tensor fields ${\mathbb T}_{ij}$ on ${\cal K}^n$ are not identically vanishing only when $n \geqslant 3$.  
The explicit form of Eq.~\eqref{eq:Teqij} on this background is of the form
\begin{align}
&r^2 \left[\:\!
F \:\!\partial_x^2+2\:\!\partial_v \partial_x+F'\:\!\partial_x
\:\!\right] H_{\mathrm{T}}
+n\:\!r\:\!r' (F\:\!\partial_x +\partial_v) H_{\mathrm{T}}
+2\:\!r(r' F'+r'' F) H_{\mathrm{T}}
\cr
&~~~~~~~~~~~~~~~~~~~~~~~~~~~~~~~~~~~~~~~~~~~~~~~~~~~~~~~~~~~~~~~~~~~
\label{eq:Teqijpde}
+2\:\!(n-1) \:\!r'^2 F H_{\mathrm{T}}
-(\mu^2 r^2+\lambda_{\mathrm{L}}) H_{\mathrm{T}}=0.
\end{align}
Let us assume that $H_{\mathrm{T}}$ is expanded as series in $\lambda$ as
\begin{align}
H_{\mathrm{T}}=\sum_{l=0}^\infty \lambda^l \cdot \Phi_{\mathrm{T}}^{(l)}. 
\end{align}
By using this expression and Eq.~(\ref{def:Delta:abcdef}), we can expand Eq.~\eqref{eq:Teqijpde} in $\lambda$
\begin{align}
\sum_{l=0}^\infty \lambda^l \cdot 
\sum_{m=0}^l \mathcal{L}_{\mathrm{T}}^{(m)} \Phi^{(l-m)}_{\mathrm{T}}=0 \,, 
\end{align} 
where $\mathcal{L}_{\mathrm{T}}^{(m)}$ is defined by
\begin{align}
\mathcal{L}_{\mathrm{T}}^{(m)}
&:=x\:\!(x+\sigma) \Delta^{(m)}_{(200;100)} \partial_x^2
+2\:\!\Delta^{(m)}_{(200;000)} \:\!\partial_v \partial_x
+n\Delta^{(m)}_{(110;000)} \partial_v
\cr
&+\left[\:\!
(2\:\!x+\sigma) \Delta^{(m)}_{(200;100)}
+x\:\!(x+\sigma) \Delta^{(m)}_{(200;010)} 
+n\:\!x\:\!(x+\sigma) \Delta^{(m)}_{(110;100)}
\:\!\right]\partial_x
+2\:\!(2\:\!x+\sigma) \Delta^{(m)}_{(110;100)}
\cr
&+2\:\!x\:\!(x+\sigma) \left[\:\!
\Delta^{(m)}_{(110;010)}+\Delta^{(m)}_{(101;100)}
+(n-1)\Delta^{(m)}_{(020;100)}
\:\!\right]-\mu^2 \Delta^{(m)}_{(200;000)}
-\delta_{m0}\:\!\lambda_{\mathrm{L}}.
\end{align} 
Since all the coefficients of $\lambda^l$ must vanish, we have
\begin{align}
\mathcal{L}_{\mathrm{T}}^{(0)} \Phi_{\mathrm{T}}^{(l)}
=-\sum_{m=1}^l \mathcal{L}_{\mathrm{T}}^{(m)}\Phi_{\mathrm{T}}^{(l-m)}\,.
\label{eq:gnrl:tensor}
\end{align} 
Thus, for the leading-order, we have the homogeneous wave equation for the single scalar variable $\Phi^{(0)}_{\rm T}$ and for higher $l$-th order, we have 
the inhomogeneous wave equation for $\Phi^{(l)}_{\rm T}$ with the source term consisting only of the lower-order variables $\Phi_{\mathrm{T}}^{(l-m)}$.

\subsection{Vector-type components}
Next we consider the vector-type components. The explicit form of Eq.~\eqref{eq:Veomij} on this background is of the form
\begin{align}
\label{eq:D2HT}
&r^2 F\:\!\partial_x^2 H_{\mathrm{T}}
+2\:\!r^2 \partial_v \partial_x H_{\mathrm{T}}
+(r^2 F'+n\:\!r\:\!r' F) \partial_x H_{\mathrm{T}}
+n\:\!r\:\!r' \partial_v H_{\mathrm{T}}
-2\:\!k_{\mathrm{V}}r'(F f_x +f_v)
\cr
&~~~~~~~~~~~~~~~~~~~~~~~~~~~~~~~~~~~~~~~~~~~
+\left[\:\!
-\mu^2 r^2 -k_{\mathrm{V}}^2-(n-1)K
+2\:\!r\:\!r'F'+2\:\!r\:\!r''F
+2\:\!(n-1)\:\!r'^2 F
\:\!\right]H_{\mathrm{T}}
=0.
\end{align}
The $a=x$ and $a=v$ components of Eq.~\eqref{eq:arrVeqa}, respectively, take the following forms: 
%
\begin{align}
&r^2 F \partial_x^2 f_x
+2\:\!r^2 \partial_x \partial_v f_x
+2\:\!r^2F' \partial_x f_x
+r\:\!r' \left[\:\!
(n+2) (F\:\!\partial_x f_x
+\partial_v f_x)
+2\:\!\partial_x f_v
\:\!\right]
+\left[\:\!
(n+2)\:\!r\:\!r''+n\:\! r'^2
\:\!\right]f_v
\cr
&~~~~~~~~~~~~\!+\left[\:\!
-\mu^2 r^2-k_{\mathrm{V}}^2 -(n-1) K+r^2 F''
+(n+4) \:\!r\:\!r' F' 
+(n+3)\:\!r\:\!r''F
+2(n-1)\:\!r'^2 F
\:\!\right] f_x
=0 \,,~~~~
\end{align}
%
\begin{align}
\label{eq:D2fv}
&r^2 F\:\!\partial_x^2 f_v
+2\:\!r^2 \partial_v \partial_x f_v
+r^2 F' \partial_v f_x
+n\:\!r\:\!r'(\partial_v f_v+F \partial_x f_v)
\cr
&~~~~~~~~~~~~~~~~~~~~~~~~~~~~~~~~~~~~~~\:\!~~~~~~
+\left[\:\!
-\mu^2 r^2-k_{\mathrm{V}}^2-(n-1)K
+2\:\!r\:\!r'F'
+(n-2)\:\!r'^2 F
+r\:\!r'' F
\:\!\right]f_v=0 \,. ~~~~
\end{align}
We expand $H_{\mathrm{H}}$ and $f_a$ as series in $\lambda$
\begin{align}
H_{\mathrm{T}}=\sum_{l=0}^\infty \lambda^l\cdot \Phi_{\mathrm{V}1}^{(l)}, 
\quad 
f_x=\sum_{l=0}^\infty \lambda^l\cdot \Phi^{(l)}_{\mathrm{V}2},
\quad 
f_v=\sum_{l=0}^\infty \lambda^l \cdot \Phi^{(l)}_{\mathrm{V}3}.
\end{align}
Expanding Eqs.~\eqref{eq:D2HT}--\eqref{eq:D2fv} in $\lambda$, together with the formula (\ref{def:Delta:abcdef}), we have
\begin{align}
\label{eq:Valphaeq}
&\sum_{l=0}^\infty \lambda^l\cdot \sum_{m=0}^l
\left[\:\!
\mathcal{L}_{\alpha 1}^{(m)} \Phi_{\mathrm{V}1}^{(l-m)}
+\mathcal{L}_{\alpha 2}^{(m)} \Phi_{\mathrm{V}2}^{(l-m)}
+\mathcal{L}_{\alpha 3}^{(m)} \Phi_{\mathrm{V}3}^{(l-m)}
\:\!\right]=0,
\\
\label{eq:Vbetaeq}
&\sum_{l=0}^\infty \lambda^l\cdot \sum_{m=0}^l
\left[\:\!
\mathcal{L}_{\beta 2}^{(m)} \Phi_{\mathrm{V}2}^{(l-m)}
+\mathcal{L}_{\beta 3}^{(m)} \Phi_{\mathrm{V}3}^{(l-m)}
\:\!\right]=0,
\\
\label{eq:Vgammaeq}
&\sum_{l=0}^\infty \lambda^l\cdot \sum_{m=0}^l
\left[\:\!
\mathcal{L}_{\gamma 2}^{(m)} \Phi_{\mathrm{V}2}^{(l-m)}
+\mathcal{L}_{\gamma 3}^{(m)} \Phi_{\mathrm{V}3}^{(l-m)}
\:\!\right]=0,
\end{align}
respectively. 
Here the differential operators $\mathcal{L}^{(m)}_{\alpha I}$ ($I=1, 2, 3$) in Eq.~\eqref{eq:Valphaeq} are defined by
\begin{align}
\mathcal{L}_{\alpha 1}^{(m)}
&:=x\:\!(x+\sigma)\Delta^{(m)}_{(200;100)}\partial_x^2
+2\:\!\Delta^{(m)}_{(200;000)} \partial_v \partial_x
+(2\:\!x+\sigma) \Delta^{(m)}_{(200;100)}\partial_x
+x\:\!(x+\sigma)\left[\:\!
\Delta^{(m)}_{(200;010)}+n\:\!\Delta^{(m)}_{(110;100)}
\:\!\right]\partial_x
\cr
&
~~~
+n\:\!\Delta^{(m)}_{(110;000)}\partial_v
+2\:\!x\:\!(x+\sigma)\left[\:\!
\Delta^{(m)}_{(110;010)}+\Delta^{(m)}_{(101;100)}
+(n-1)\Delta^{(m)}_{(020;100)}\:\!\right]
+2\:\!(2\:\!x+\sigma)\Delta^{(m)}_{(110;100)}
\cr
&~~~
-\mu^2\:\!\Delta^{(m)}_{(200;000)}
-\delta_{m0}\left[\:\!k_\mathrm{V}^2+(n-1)K\:\!\right],
\\
\mathcal{L}_{\alpha2}^{(m)}
&:=-2\:\!k_\mathrm{V} x\:\!(x+\sigma)\Delta^{(m)}_{(010;100)},
\\
\mathcal{L}_{\alpha 3}^{(m)}
&:=-2\:\!k_{\mathrm{V}}\Delta^{(m)}_{(010;000)},
\end{align}
the differential operators $\mathcal{L}^{(m)}_{\beta I}$ ($I=2, 3$) in Eq.~\eqref{eq:Vbetaeq} are defined by
\begin{align}
\mathcal{L}_{\beta 2}^{(m)}
&:=
x\:\!(x+\sigma)\Delta^{(m)}_{(200;100)}\partial_x^2
+2\:\!\Delta^{(m)}_{(200;000)}\partial_x\partial_v
+(n+2) \Delta^{(m)}_{(110;000)}\partial_v
+2\:\!(2\:\!x+\sigma) \Delta^{(m)}_{(200;100)}\partial_x
\cr
&+x\:\!(x+\sigma) \left[\:\!
2\:\!\Delta^{(m)}_{(200;010)}+(n+2) \Delta^{(m)}_{(110;100)}
\:\!\right]\partial_x
+(2\:\!x+\sigma)\left[\:\!
2\:\!\Delta^{(m)}_{(200;010)}
+(n+4)\Delta^{(m)}_{(110;100)}
\:\!\right]
\cr
&+x\:\!(x+\sigma)\left[\:\!
(n+4) \Delta^{(m)}_{(110;010)}
+(n+3) \Delta^{(m)}_{(101;100)}
+2(n-1)\Delta^{(m)}_{(020;100)}
+\Delta^{(m)}_{(200;001)}
\:\!\right]
\cr
&
+2\:\!\Delta^{(m)}_{(200;100)}
-\mu^2\:\!\Delta^{(m)}_{(200;000)}
-\delta_{m0}\left[\:\!
k_{\mathrm{V}}^2+(n-1)K
\:\!\right],
\\[3mm]
\mathcal{L}_{\beta 3}^{(m)}
&:=2\:\!\Delta^{(m)}_{(110;000)}\partial_x
+(n+2) \Delta^{(m)}_{(101;000)}
+n\:\!\Delta_{(020;000)}^{(m)},
\end{align}
and the differential operators $\mathcal{L}^{(m)}_{\gamma I}$ ($I=2, 3$) in Eq.~\eqref{eq:Vgammaeq} are defined by
\begin{align}
\mathcal{L}_{\gamma2}^{(m)}
&:=\left[\:\!
(2\:\!x+\sigma) \Delta^{(m)}_{(200;100)}
+x\:\!(x+\sigma) \Delta^{(m)}_{(200;010)}\:\!\right] \partial_v,
\\[3mm]
\mathcal{L}_{\gamma3}^{(m)}
&:=
x\:\!(x+\sigma) \Delta^{(m)}_{(200;100)} \partial_x^2
+2\:\!\Delta^{(m)}_{(200;000)} \partial_v \partial_x
+n\:\!\Delta^{(m)}_{(110;000)} \partial_v
+n\:\!x\:\!(x+\sigma)\Delta^{(m)}_{(110;100)} \partial_x
\cr
&
+x\:\!(x+\sigma) \left[\:\!
2\:\!\Delta^{(m)}_{(110;010)}
+(n-2)\Delta^{(m)}_{(020;100)}
+\Delta^{(m)}_{(101;100)}
\:\!\right]
+2\:\!(2\:\!x+\sigma)\Delta^{(m)}_{(110;100)}
\cr
&-\mu^2\:\!\Delta^{(m)}_{(200;000)}
-\delta_{m0}\left[\:\!
k_{\mathrm{V}}^2+(n-1)K
\:\!\right].
\end{align}
Since all the coefficients of $\lambda^l$ must vanish, we obtain the equations for $\Phi^{(l)}_{\mathrm{V} I}$ ($I=1, 2, 3$)
\begin{align}
\left[\:\!
\begin{array}{ccc}
\mathcal{L}^{(0)}_{\alpha1}&0&0\\[1mm]
0&\mathcal{L}^{(0)}_{\beta2}&0\\[1mm]
0&\mathcal{L}^{(0)}_{\gamma2}&\mathcal{L}^{(0)}_{\gamma3}
\end{array}
\:\!\right]\left[\:\!
\begin{array}{c}
\Phi_{\mathrm{V}1}^{(l)}\\[1mm]
\Phi_{\mathrm{V}2}^{(l)}\\[1mm]
\Phi_{\mathrm{V}3}^{(l)}
\end{array}
\:\!\right]
=-\sum_{m=1}^l
\left[\:\!
\begin{array}{ccc}
\mathcal{L}^{(m)}_{\alpha1}&\mathcal{L}^{(m)}_{\alpha2}&\mathcal{L}^{(m)}_{\alpha3}\\[1mm]
0&\mathcal{L}^{(m)}_{\beta2}&\mathcal{L}^{(m)}_{\beta3}\\[1mm]
0&\mathcal{L}^{(m)}_{\gamma2}&\mathcal{L}^{(m)}_{\gamma3}
\end{array}
\:\!\right]\left[\:\!
\begin{array}{c}
\Phi_{\mathrm{V}1}^{(l-m)}\\[1mm]
\Phi_{\mathrm{V}2}^{(l-m)}\\[1mm]
\Phi_{\mathrm{V}3}^{(l-m)}
\end{array}
\:\!\right],
\end{align}
where we have used that $\mathcal{L}^{(0)}_{\alpha2}$, $\mathcal{L}^{(0)}_{\alpha3}$, 
$\mathcal{L}^{(0)}_{\beta3}$ vanish because 
the background reduces to the near-horizon geometry at the leading-order of $\lambda$-series. 
We note that as in the Reissner--Nordstrom case, the equations for the two scalar variables $(\Phi^{(l)}_{\mathrm{V} 1}, \Phi^{(l)}_{\mathrm{V} 2})$ are mutually decoupled wave equations with the source terms consisting only of the lower-order variables. Thus, once we have solved the leading-order 
homogeneous wave equations, we can successively solve the equations above at, in principle, all order.

\subsection{Scalar-type components}
Let us finally consider the scalar-type components. First we find that each component of Eq.~\eqref{eq:Seqabarr} takes the following form: 
%
\begin{align}
\label{eq:Sxxgen}
&r^2 F \partial_x^2 f_{xx}+2\:\!r^2 \partial_v \partial_x f_{xx}
+(3\:\!r^2 F'+n\:\!r\:\!r' F) \partial_x f_{xx}
+n\:\!r\:\!r'\partial_v f_{xx}
+2(n+2)(
rr''-r'^2 
)f_{xv}
-4\:\!k_{\textrm{S}}\:\!r' f_x
\cr
&~~~~~~~~~~~~~~~~~~~~~~~~~~~~~\:\!~~
+\left[\:\!
-\mu^2 r^2-k_{\textrm{S}}^2+3\:\!r^2 F''+2\:\!n\:\! r\:\!r' F'-2\:\!(n+1) \:\!r'^2 F
+2\:\!(n+1)\:\!r\:\!r''F
\:\!\right] f_{xx}
=0 \,,~~~~~
\end{align}
%
%
\begin{align}
&r^2 F \partial_x^2 f_{xv}
+2\:\!r^2 \partial_v \partial_x f_{xv}
+r^2 F' \partial_v f_{xx}
+r^2F' \partial_x f_{xv}
+n\:\!r\:\!r'(F\:\!\partial_x f_{xv}+\partial_v f_{xv})
+\left[\:\!
\frac{r^2 F'^2}{2}-r^2 FF''+r\:\!r'FF'
\:\!\right]f_{xx}
\cr
&~~~~~~~~~~~~\:
+\left[\:\!
-\mu^2r^2-k_{\textrm{S}}^2
+(n+2) \:\!r\:\!r' F'-n\:\!r'^2 F+n\:\!r\:\!r'' F
\:\!\right]f_{xv} 
+n\:\!(r\:\!r''-r'^2) f_{vv}
-2\:\!k_{\textrm{S}}\:\!r' f_v=0 \,, 
\end{align}
%
\begin{align}
&r^2 F \partial_x^2 f_{vv}
+2\:\!r^2 \partial_x \partial_v f_{vv}
+r^2 F'(2\:\!\partial_v f_{xv}-\partial_x f_{vv})
+n\:\!r\:\!r'(F\:\!\partial_x f_{vv}+\partial_v f_{vv})
+\left[\:\!
-\mu^2 r^2 -k_{\textrm{S}}^2 +r^2 F''
\:\!\right]f_{vv}
\cr
&~~~~~~~~~~~~~~~~~~~~~~~~~~~~~~~
+\left[\:\!
r^2 F^2 F''
-\frac{r^2 FF'^2}{2}
-r\:\!r'F^2F'
\:\!\right]f_{xx}
+\left[\:\!
2\:\!r^2 FF''-r^2 F'^2
-2r\:\!r'FF'
\:\!\right]f_{xv}=0 \,. 
\end{align}
The above equations are, respectively, the $(a, b)=(x,x)$-, $(x,v)$-, and $(a,b)=(v,v)$-components of Eq.~\eqref{eq:Seqabarr}. 

\par

Next, we find that each component of Eq.~\eqref{eq:Seqaiarr} takes the following form: 
%
\begin{align}
&r^2 F \partial_x^2 f_x
+2\:\!r^2\partial_v \partial_x f_x
+\left[\:\!
2\:\!r^2 F'+(n+2) \:\!r\:\!r' F
\:\!\right]\partial_x f_x
+(n+2)\:\!r\:\!r'\partial_v f_x
+2\:\!r\:\!r'\partial_x f_v
-2\:\!k_{\textrm{S}}\:\!r'(F f_{xx}+f_{vx})
\cr
&
+\left[\:\!
n\:\!r'^2+(n+2)\:\!r\:\!r''
\:\!\right]f_v
+\left[\:\!-\mu^2 r^2-k_{\textrm{S}}^2
+r^2 F''+(n+4)\:\!r \:\!r' F'
+2\:\!(n-1)\:\!r'^2 F
+(n+3)\:\!r\:\!r'' F
\:\!\right]f_x
=0 \,,~~~~
\end{align}
%
\begin{align}
&r^2 F \:\!\partial_x^2 f_v
+2 \:\!r^2 \partial_x \partial_v f_v
+r^2 F' \partial_v f_x
+n\:\!r\:\!r' \partial_v f_v
+n\:\!r\:\!r' F\:\!\partial_x f_v
-2\:\!k_{\textrm{S}}\:\!r'(Ff_{xv}+f_{vv})
\cr
&~~~~~~~~~~~~~~~~~~~~~~~\:~~~~~~~~~~~~~~~~~~~~~~~~~~~~~~~~~~~~
+\left[\:\!
-\mu^2 r^2-k_{\textrm{S}}^2+2\:\!r\:\!r' F' 
+r\:\!r'' F
+(n-2)\:\!r'^2 F
\:\!\right]f_v
=0 \,.~~~~~
\end{align}
These two equations are, respectively, $a=x$- and $a=v$-component of Eq.~\eqref{eq:Seqaiarr}. 

\par 

Finally, we find that the explicit forms of Eqs.~\eqref{eq:Seqijtless} and \eqref{eq:Seqijt} are 
\begin{align}
&r^2\left[\:\!
F\:\!\partial_x^2+2\partial_v\partial_x+F' \partial_x
\:\!\right]H_{\mathrm{T}}
+n\:\!r\:\!r' (F\:\!\partial_x +\partial_v) H_{\mathrm{T}}
-2\:\!k\:\!r'(F f_x+f_v)
\cr
&~~~~~~~~~~~~~~~~~~~~~~~~~~~~~~~\:\!~~~~~~~~~~~~~~~~~~~~~~~
+\left[\:\!
-\mu^2 r^2-k_{\textrm{S}}^2+2\:\!(n-1)\:\! r'^2 F+2\:\!r(r'F'+r'' F)
\:\!\right]H_{\mathrm{T}}
=0 \,, 
\\
\label{eq:SHLgen}
&r^2 (F\:\!\partial_x^2+2\partial_v \partial_x +F' \partial_x)H_{\mathrm{L}}
+n\:\!r\:\!r' (F\:\!\partial_x+\partial_v)H_{\mathrm{L}}
+r'^2 (F^2 f_{xx}+2Ff_{xv}+f_{vv})
-r\:\!r'F' \left[\:\!
\frac{F f_{xx}}{2} +f_{xv}
\:\!\right]
\cr
&~~~~~~~~~~
-r\:\!r''(F^2 f_{xx}+2F f_{xv}+f_{vv})
+2\frac{k_{\textrm{S}}}{n} r' (F f_x +f_v)
+\left[\:\!
-\mu^2 r^2-k_{\textrm{S}}^2-2\:\!r'^2 F+2\:\!r(r' F'+r'' F)
\:\!\right]H_{\mathrm{L}}=0 \,. ~~~~~
\end{align}

\par 

Now we assume that we can expand $f_{ab}$, $f_a$, $H_{\mathrm{L}}$, and $ H_{\mathrm{T}}$ as series in $\lambda$
\begin{align}
&
H_{\mathrm{T}}=\sum_{l=0}^\infty \lambda^l \cdot \Phi_{\mathrm{S}1}^{(l)}
\quad
H_{\mathrm{L}}=\sum_{l=0}^\infty \lambda^l \cdot \Phi_{\mathrm{S}2}^{(l)},
\quad
f_{x}=\sum_{l=0}^\infty \lambda^l \cdot \Phi_{\mathrm{S}3}^{(l)},
\quad
f_{xx}=\sum_{l=0}^\infty \lambda^l \cdot \Phi_{\mathrm{S}4}^{(l)},
\\
&
f_{v}=\sum_{l=0}^\infty \lambda^l \cdot \Phi_{\mathrm{S}5}^{(l)},
\quad
f_{xv}=\sum_{l=0}^\infty \lambda^l \cdot \Phi_{\mathrm{S}6}^{(l)},
\quad
f_{vv}=\sum_{l=0}^\infty \lambda^l \cdot \Phi_{\mathrm{S}7}^{(l)}.
\quad
\end{align}
Expanding Eqs.~\eqref{eq:Sxxgen}--\eqref{eq:SHLgen} in $\lambda$ and using the formula (\ref{def:Delta:abcdef}), we obtain
\begin{align}
\label{eq:Sdelta}
&\sum_{l=0}^\infty\lambda^l\cdot 
\sum_{m=0}^l \left[\:\!
\mathcal{L}_{\delta 3}^{(m)} \Phi_{\mathrm{S}3}^{(l-m)}
+\mathcal{L}_{\delta 4}^{(m)} \Phi_{\mathrm{S}4}^{(l-m)}
+\mathcal{L}_{\delta 6}^{(m)} \Phi_{\mathrm{S}6}^{(l-m)}
\:\!\right]=0,
\\
\label{eq:Szeta}
&\sum_{l=0}^\infty\lambda^l\cdot 
\sum_{m=0}^l \left[\:\!
\mathcal{L}_{\zeta 4}^{(m)} \Phi_{\mathrm{S}4}^{(l-m)}
+\mathcal{L}_{\zeta 5}^{(m)} \Phi_{\mathrm{S}5}^{(l-m)}
+\mathcal{L}_{\zeta 6}^{(m)} \Phi_{\mathrm{S}6}^{(l-m)}
+\mathcal{L}_{\zeta 7}^{(m)} \Phi_{\mathrm{S}7}^{(l-m)}
\:\!\right]=0,
\\
\label{eq:Seta}
&\sum_{l=0}^\infty\lambda^l\cdot 
\sum_{m=0}^l \left[\:\!
\mathcal{L}_{\eta 4}^{(m)} \Phi_{\mathrm{S}4}^{(l-m)}
+\mathcal{L}_{\eta 6}^{(m)} \Phi_{\mathrm{S}6}^{(l-m)}
+\mathcal{L}_{\eta 7}^{(m)} \Phi_{\mathrm{S}7}^{(l-m)}
\:\!\right]=0 \,,
\\
\label{eq:Sgamma}
&\sum_{l=0}^\infty\lambda^l\cdot 
\sum_{m=0}^l \left[\:\!
\mathcal{L}_{\gamma 3}^{(m)} \Phi_{\mathrm{S}3}^{(l-m)}
+\mathcal{L}_{\gamma 4}^{(m)} \Phi_{\mathrm{S}4}^{(l-m)}
+\mathcal{L}_{\gamma 5}^{(m)} \Phi_{\mathrm{S}5}^{(l-m)}
+\mathcal{L}_{\gamma 6}^{(m)} \Phi_{\mathrm{S}6}^{(l-m)}
\:\!\right]=0,
\\
\label{eq:Sepsilon}
&\sum_{l=0}^\infty\lambda^l\cdot 
\sum_{m=0}^l \left[\:\!
\mathcal{L}_{\epsilon 3}^{(m)} \Phi_{\mathrm{S}3}^{(l-m)}
+\mathcal{L}_{\epsilon 5}^{(m)} \Phi_{\mathrm{S}5}^{(l-m)}
+\mathcal{L}_{\epsilon 6}^{(m)} \Phi_{\mathrm{S}6}^{(l-m)}
+\mathcal{L}_{\epsilon 7}^{(m)} \Phi_{\mathrm{S}7}^{(l-m)}
\:\!\right]=0,
\\
\label{eq:Salpha}
&\sum_{l=0}^\infty\lambda^l\cdot 
\sum_{m=0}^l \left[\:\!
\mathcal{L}_{\alpha 1}^{(m)} \Phi_{\mathrm{S}1}^{(l-m)}
+\mathcal{L}_{\alpha 3}^{(m)} \Phi_{\mathrm{S}3}^{(l-m)}
+\mathcal{L}_{\alpha 5}^{(m)} \Phi_{\mathrm{S}5}^{(l-m)}
\:\!\right]=0,
\\
\label{eq:Sbeta}
&\sum_{l=0}^\infty\lambda^l\cdot 
\sum_{m=0}^l \left[\:\!
\mathcal{L}_{\beta 2}^{(m)} \Phi_{\mathrm{S}2}^{(l-m)}
+\mathcal{L}_{\beta 3}^{(m)} \Phi_{\mathrm{S}3}^{(l-m)}
+\mathcal{L}_{\beta 4}^{(m)} \Phi_{\mathrm{S}4}^{(l-m)}
+\mathcal{L}_{\beta 5}^{(m)} \Phi_{\mathrm{S}5}^{(l-m)}
+\mathcal{L}_{\beta 6}^{(m)} \Phi_{\mathrm{S}6}^{(l-m)}
+\mathcal{L}_{\beta 7}^{(m)} \Phi_{\mathrm{S}7}^{(l-m)}
\:\!\right]=0.
\end{align}
Here, the differential operators $\mathcal{L}^{(m)}_{\alpha I}$ ($I=1, 3, 5$) in Eq.~\eqref{eq:Salpha} are defined by
\begin{align}
\mathcal{L}_{\alpha1}^{(m)}
&:=x(x+\sigma )\Delta^{(m)}_{(200;100)}\partial_x^2
+2\:\!\Delta^{(m)}_{(200;000)} \partial_v \partial_x
+n\:\!\Delta^{(m)}_{(110;000)}\partial_v
+(2\:\!x+\sigma)\Delta^{(m)}_{(200;100)} \partial_x
\cr
&~~~~
+x\:\!(x+\sigma) \left[\:\!
\Delta^{(m)}_{(200;010)}+n\:\!\Delta^{(m)}_{(110;100)}
\:\!\right]\partial_x 
+2\:\!x\:\!(x+\sigma) \left[\:\!
(n-1)\Delta^{(m)}_{(020;100)}
+\Delta^{(m)}_{(110;010)}
+\Delta^{(m)}_{(101;100)}
\:\!\right]
\cr
&~~~~
+2\:\!(2\:\!x+\sigma) \Delta^{(m)}_{(110;100)}
-\mu^2\:\!\Delta^{(m)}_{(200;000)}
-\delta_{m0} k_{\textrm{S}}^2,
\\
\mathcal{L}^{(m)}_{\alpha 3}
&:=-2\:\!k_{\textrm{S}}\:\!x\:\!(x+\sigma)\Delta^{(m)}_{(010;100)},
\\
\mathcal{L}^{(m)}_{\alpha 5}
 &:=-2\:\!k_{\textrm{S}}\:\!\Delta^{(m)}_{(010;000)},
\end{align}
the differential operators $ \mathcal{L}_{\beta I}$ ($I=2,\ldots, 7$) in Eq.~\eqref{eq:Sbeta} are given by
\begin{align}
\mathcal{L}_{\beta 2}^{(m)}
&:=x(x+\sigma )\Delta^{(m)}_{(200;100)}\partial_x^2
+2\:\!\Delta^{(m)}_{(200;000)} \partial_v \partial_x
+n\:\!\Delta^{(m)}_{(110;000)} \partial_v
+(2\:\!x+\sigma)\Delta^{(m)}_{(200;100)} \partial_x
\cr
&~~~~
+x\:\!(x+\sigma) \left[\:\!
\Delta^{(m)}_{(200;010)} 
+n\:\!\Delta^{(m)}_{(110;100)}
\:\!\right]\partial_x
+2\:\!x\:\!(x+\sigma)\left[\:\!
\Delta^{(m)}_{(110;010)}
-\Delta^{(m)}_{(020;100)}
+\Delta^{(m)}_{(101;100)}
\:\!\right]
\cr
&~~~~
+2(2\:\!x+\sigma)\Delta^{(m)}_{(110;100)}
-\mu^2\:\!\Delta^{(m)}_{(200;000)}
-\delta_{m0}k_\textrm{S}^2,
\\
\mathcal{L}_{\beta 3}^{(m)}
 &:=\frac{2\:\!k_{\textrm{S}}}{n} x\:\!(x+\sigma) \Delta^{(m)}_{(010;100)},
\\
\mathcal{L}_{\beta 4}^{(m)}
&:=x^2(x+\sigma)^2\left[\:\!
\Delta^{(m)}_{(020;200)}
-\frac{1}{2} \Delta^{(m)}_{(110;110)}
- \Delta^{(m)}_{(101;200)}
\:\!\right]
-\frac{x\:\!(x+\sigma)(2\:\!x+\sigma)}{2} \Delta^{(m)}_{(110;200)},
\\
\mathcal{L}_{\beta 5}^{(m)}
&:=\frac{2\:\!k_{\textrm{S}}}{n} \Delta^{(m)}_{(010;000)},
\\
\mathcal{L}_{\beta 6}^{(m)}
&:=x\:\!(x+\sigma)\left[\:\!
2\:\!\Delta^{(m)}_{(020;100)}
-2\:\!\Delta^{(m)}_{(101;100)}
-\Delta^{(m)}_{(110;010)}
\:\!\right]
-(2\:\!x+\sigma) \Delta^{(m)}_{(110;100)},
\\
\mathcal{L}_{\beta 7}^{(m)}
&:=\Delta^{(m)}_{(020;000)}
-\Delta^{(m)}_{(101;000)},
\end{align}
the differential operators $ \mathcal{L}^{(m)}_{\gamma I}$ ($I=3, 4, 5, 6$) in Eq.~\eqref{eq:Sgamma} are defined by
\begin{align}
\mathcal{L}_{\gamma 3}^{(m)}
&:=x\:\!(x+\sigma)\Delta^{(m)}_{(200;100)}\partial_x^2
+2\:\!\Delta^{(m)}_{(200;000)} \partial_v \partial_x
+(n+2)\Delta^{(m)}_{(110;000)} \partial_v
+2\:\!(2\:\!x+\sigma)\Delta^{(m)}_{(200;100)} \partial_x
\cr
&~~~~
+x\:\!(x+\sigma) \left[\:\!
2\:\!\Delta^{(m)}_{(200;010)}+(n+2)\Delta^{(m)}_{(110;100)}
\:\!\right]\partial_x
+(2\:\!x+\sigma) \left[\:\!
2\:\!\Delta^{(m)}_{(200;010)}
+(n+4) \Delta^{(m)}_{(110;100)}
\:\!\right]
\cr
&~~~~
+x\:\!(x+\sigma) \left[\:\!
\Delta^{(m)}_{(200;001)}
+(n+4)\Delta^{(m)}_{(110;010)}
+2(n-1)\Delta^{(m)}_{(020;100)}
+(n+3)\Delta^{(m)}_{(101;100)}
\:\!\right]
\cr
&~~~~
+2\:\!\Delta^{(m)}_{(200;100)}
-\mu^2\:\!\Delta^{(m)}_{(200;000)}-\delta_{m0} k_{\textrm{S}}^2,
\\
\mathcal{L}_{\gamma 4}^{(m)}
&:=-2\:\!k_{\textrm{S}}\:\!x\:\!(x+\sigma) \Delta^{(m)}_{(010;100)},
\\
\mathcal{L}^{(m)}_{\gamma 5}
&:=2\:\!\Delta^{(m)}_{(110;000)}\partial_x
+n\:\!\Delta^{(m)}_{(020;000)}
+(n+2)\Delta^{(m)}_{(101;000)},
\\
\mathcal{L}_{\gamma 6}^{(m)}
&:= -2\:\!k_{\textrm{S}}\:\!\Delta^{(m)}_{(010;000)},
\end{align}
the differential operator $\mathcal{L}_{\delta I}^{(m)}$ ($I=3, 4, 6$) in Eq.~\eqref{eq:Sdelta} are defined by 
\begin{align}
\mathcal{L}_{\delta 3}^{(m)}
&:=-4\:\!k_{\textrm{S}}\:\!\Delta^{(m)}_{(010;000)},
\\
\mathcal{L}_{\delta4}^{(m)}
&:=x\:\!(x+\sigma) \Delta^{(m)}_{(200;100)}\partial_x^2
+2\:\!\Delta^{(m)}_{(200;000)} \partial_v \partial_x
+n\:\!\Delta^{(m)}_{(110;000)} \partial_v
+3(2\:\!x+\sigma) \Delta^{(m)}_{(200;100)} \partial_x
\cr
&~~~~+x\:\!(x+\sigma)\left[\:\!
3\:\!\Delta^{(m)}_{(200;010)}+n\:\!\Delta^{(m)}_{(110;100)}
\:\!\right] \partial_x
+(2\:\!x+\sigma)\left[\:\!
6\Delta^{(m)}_{(200;010)}+2\:\!n\:\!\Delta^{(m)}_{(110;100)}
\:\!\right]
\cr
&
~~~~+x\:\!(x+\sigma)\left[\:\!
3\:\!\Delta^{(m)}_{(200;001)}
+2\:\!n\:\!\Delta^{(m)}_{(110;010)}
-2(n+1) \Delta^{(m)}_{(020;100)}
+2(n+1) \Delta^{(m)}_{(101;100)}
\:\!\right]
\cr
&~~~~
-\mu^2\:\!\Delta^{(m)}_{(200;000)}
-\delta_{m0}k_{\textrm{S}}^2
+6\Delta^{(m)}_{(200;100)},
\\
\mathcal{L}_{\delta 6}^{(m)}
&:=2(n+2)\left[\:\!
\Delta^{(m)}_{(101;000)}-\Delta^{(m)}_{(020;000)}
\:\!\right],
\end{align}
the differential operators $ \mathcal{L}^{(m)}_{\epsilon I}$ ($I=3, 5, 6, 7$) in Eq.~\eqref{eq:Sepsilon} are defined by
\begin{align}
\mathcal{L}_{\epsilon3}^{(m)}
&:=\left[\:\!
(2\:\!x+\sigma)\Delta^{(m)}_{(200;100)}
+x\:\!(x+\sigma)\Delta^{(m)}_{(200;100)}
\:\!\right]\partial_v,
\\
\mathcal{L}_{\epsilon 5}^{(m)}
&:=x\:\!(x+\sigma) \Delta^{(m)}_{(200;100)} \partial_x^2
+2\:\!\Delta^{(m)}_{(200;000)} \partial_x \partial_v
+n\:\!\Delta^{(m)}_{(110;000)}\partial_v
+n\:\!x\:\!(x+\sigma)\Delta^{(m)}_{(110;100)} \partial_x
+2 (2\:\!x+\sigma) \Delta^{(m)}_{(110;100)}
\cr
&~~~~
+x\:\!(x+\sigma)\left[\:\!
2\:\!\Delta^{(m)}_{(110;010)}
+
(n-2)\Delta^{(m)}_{(020;100)}
+\Delta^{(m)}_{(101;100)}
\:\!\right]
-\mu^2\:\!\Delta^{(m)}_{(200;000)}
-\delta_{m0} k_{\textrm{S}}^2,
\\
\mathcal{L}_{\epsilon 6}^{(m)}
&:=-\:\!2\:\!k_{\textrm{S}}\:\! x\:\!(x+\sigma) \Delta^{(m)}_{(010;100)},
\\
\mathcal{L}_{\epsilon 7}^{(m)}
&:=-2\:\!k_{\textrm{S}}\:\! \Delta^{(m)}_{(010;000)},
\end{align}
the differential operators $\mathcal{L}_{\zeta I}^{(m)}$ ($I=4,5,6,7$) in Eq.~\eqref{eq:Szeta} are defined by
\begin{align}
\mathcal{L}_{\zeta4}^{(m)}
&:=x\:\!(x+\sigma) \Delta^{(m)}_{(200;010)}
+(2\:\!x+\sigma)\Delta^{(m)}_{(200;100)}
+\frac{\sigma^2}{2}\Delta^{(m)}_{(200;200)}
+x\:\!(x+\sigma)(2\:\!x+\sigma)\left[\:\!
\Delta^{(m)}_{(110;200)}
-\Delta^{(m)}_{(200;110)}
\:\!\right]
\cr
&~~~~
+x^2(x+\sigma)^2\left[\:\!
\frac{1}{2}\Delta^{(m)}_{(200;020)}
-\Delta^{(m)}_{(200;101)}
+\Delta^{(m)}_{(110;110)}
\:\!\right],
\\
\mathcal{L}_{\zeta5}^{(m)}
&:=-2\:\!k_{\textrm{S}}\:\!\Delta^{(m)}_{(010;000)}, 
\\
\mathcal{L}_{\zeta6}^{(m)}
&:=x\:\!(x+\sigma) \Delta^{(m)}_{(200;100)} \partial_x^2
+2\:\!\Delta^{(m)}_{(200;000)} \partial_v \partial_x
+n\:\!\Delta^{(m)}_{(110;000)} \partial_v
\cr
&~~~~
+
(2\:\!x+\sigma)\Delta^{(m)}_{(200;100)}\partial_x
+x\:\!(x+\sigma)\left[\:\!
n\:\! \Delta^{(m)}_{(110;100)}
+\Delta^{(m)}_{(200;010)}\:\!\right]\partial_x
+(n+2)(2\:\!x+\sigma)\Delta^{(m)}_{(110;100)}
\cr
&~~~~
+x\:\!(x+\sigma)\left[\:\!
(n+2)\Delta^{(m)}_{(110;010)}
-n\:\!\Delta^{(m)}_{(020;100)}
+n\:\!\Delta^{(m)}_{(101;100)}
\:\!\right]
-\mu^2\:\!\Delta^{(m)}_{(200;000)}
-\delta_{m0}k_{\textrm{S}}^2,
\\
\mathcal{L}_{\zeta 7}^{(m)}
&:=n\left[\:\!
\Delta^{(m)}_{(101;000)}-\Delta^{(m)}_{(020;000)}
\:\!\right],
\end{align}
and the differential operators $\mathcal{L}_{\eta I}^{(m)}$ ($I=4, 6, 7$) in Eq.~\eqref{eq:Seta} are defined by
\begin{align}
\mathcal{L}_{\eta 4}^{(m)}
&:=-\frac{\sigma^2}{2}x\:\!(x+\sigma)  \Delta^{(m)}_{(200;300)}
+x^2(x+\sigma)^2(2\:\!x+\sigma) 
\left[ \Delta^{(m)}_{(200;210)}-\Delta^{(m)}_{(110;300)} \right] 
\cr
&~~~~
+x^3(x+\sigma)^3\left[\:\!
\Delta^{(m)}_{(200;201)}-\frac{1}{2}\Delta^{(m)}_{(200;120)}
-\Delta^{(m)}_{(110;210)}
\:\!\right],
\\
\mathcal{L}_{\eta 6}^{(m)}
&:=
2\left[\:\!
(2\:\!x+\sigma) \Delta^{(m)}_{(200;100)}+x\:\!(x+\sigma) \Delta^{(m)}_{(200;010)}
\:\!\right]\partial_v
+2\:\!x\:\!(x+\sigma)(2\:\!x+\sigma)\left[\:\!
\Delta^{(m)}_{(200;110)}-\Delta^{(m)}_{(110;200)}
\:\!\right]
\cr
&~~~~
+x^2(x+\sigma)^2\left[\:\!
2\:\!\Delta^{(m)}_{(200;101)}
-\Delta^{(m)}_{(200;020)}
-2\:\!\Delta^{(m)}_{(110;110)}
\:\!\right]
-\sigma^2\:\!\Delta^{(m)}_{(200;200)},
\\
\mathcal{L}_{\eta 7}^{(m)}
&:=x\:\!(x+\sigma)\Delta^{(m)}_{(200;100)} \partial_x^2
+2\:\!\Delta^{(m)}_{(200;000)} \partial_x \partial_v
+n\:\!\Delta^{(m)}_{(110;000)} \partial_v
-(2\:\!x+\sigma)\Delta^{(m)}_{(200;100)} \partial_x
\cr
&~~~~
+x\:\!(x+\sigma)\left[\:\!
n\:\!\Delta^{(m)}_{(110;100)}-\Delta^{(m)}_{(200;010)}
\:\!\right]\partial_x
+2(2\:\!x+\sigma)
\Delta^{(m)}_{(200;010)}
+x\:\!(x+\sigma) \Delta^{(m)}_{(200;001)}
\cr
&~~~~
+2\:\!\Delta^{(m)}_{(200;100)}
-\mu^2\:\!\Delta^{(m)}_{(200;000)}
-\delta_{m0} k_{\textrm{S}}^2.
\end{align}
Since all the coefficients of $\lambda^l$ must vanish, we have 
\begin{align}
\left[\:\!
\begin{array}{ccccccc}
\mathcal{L}_{\alpha 1}^{(0)}&
0&
0&
0&
0&
0&
0\\[1mm]
0&
\mathcal{L}_{\beta 2}^{(0)}&
0&
0&
0&
0&
0\\[1mm]
0&
0&
\mathcal{L}_{\gamma 3}^{(0)}&
0&
0&
0&
0\\[1mm]
0&
0&
0&
\mathcal{L}_{\delta 4}^{(0)}&
0&
0&
0\\[1mm]
0&
0&
\mathcal{L}_{\epsilon 3}^{(0)}&
0&
\mathcal{L}_{\epsilon 5}^{(0)}&
0&
0\\[1mm]
0&
0&
0&
\mathcal{L}_{\zeta 4}^{(0)}&
0&
\mathcal{L}_{\zeta 6}^{(0)}&
0\\[1mm]
0&
0&
0&
\mathcal{L}_{\eta 4}^{(0)}&
0&
\mathcal{L}_{\eta 6}^{(0)}&
\mathcal{L}_{\eta 7}^{(0)}\\
\end{array}
\:\!\right]\left[\:\!
\begin{array}{c}
\Phi^{(l)}_{\mathrm{S}1}\\[1mm]
\Phi^{(l)}_{\mathrm{S}2}\\[1mm]
\Phi^{(l)}_{\mathrm{S}3}\\[1mm]
\Phi^{(l)}_{\mathrm{S}4}\\[1mm]
\Phi^{(l)}_{\mathrm{S}5}\\[1mm]
\Phi^{(l)}_{\mathrm{S}6}\\[1mm]
\Phi^{(l)}_{\mathrm{S}7}\\[1mm]
\end{array}
\:\!\right]
=-\sum_{m=1}^l
\left[\:\!
\begin{array}{ccccccc}
\mathcal{L}_{\alpha 1}^{(m)}&
0&
\mathcal{L}_{\alpha 3}^{(m)}&
0&
\mathcal{L}_{\alpha 5}^{(m)}&
0&
0\\[1mm]
0&
\mathcal{L}_{\beta 2}^{(m)}&
\mathcal{L}_{\beta 3}^{(m)}&
\mathcal{L}_{\beta 4}^{(m)}&
\mathcal{L}_{\beta 5}^{(m)}&
\mathcal{L}_{\beta 6}^{(m)}&
\mathcal{L}_{\beta 7}^{(m)}
\\[1mm]
0&
0&
\mathcal{L}_{\gamma 3}^{(m)}&
\mathcal{L}_{\gamma 4}^{(m)}&
\mathcal{L}_{\gamma 5}^{(m)}&
\mathcal{L}_{\gamma 6}^{(m)}&
0\\[1mm]
0&
0&
\mathcal{L}_{\delta 3}^{(m)}&
\mathcal{L}_{\delta 4}^{(m)}&
0&
\mathcal{L}_{\delta 6}^{(m)}&
0\\[1mm]
0&
0&
\mathcal{L}_{\epsilon 3}^{(m)}&
0&
\mathcal{L}_{\epsilon 5}^{(m)}&
\mathcal{L}_{\epsilon 6}^{(m)}&
\mathcal{L}_{\epsilon 7}^{(m)}
\\[1mm]
0&
0&
0&
\mathcal{L}_{\zeta 4}^{(m)}&
\mathcal{L}_{\zeta 5}^{(m)}&
\mathcal{L}_{\zeta 6}^{(m)}&
\mathcal{L}_{\zeta 7}^{(m)}
\\[1mm]
0&
0&
0&
\mathcal{L}_{\eta 4}^{(m)}&
0&
\mathcal{L}_{\eta 6}^{(m)}&
\mathcal{L}_{\eta 7}^{(m)}\\[1mm]
\end{array}
\:\!\right]\left[\:\!
\begin{array}{c}
\Phi^{(l-m)}_{\mathrm{S}1}\\[1mm]
\Phi^{(l-m)}_{\mathrm{S}2}\\[1mm]
\Phi^{(l-m)}_{\mathrm{S}3}\\[1mm]
\Phi^{(l-m)}_{\mathrm{S}4}\\[1mm]
\Phi^{(l-m)}_{\mathrm{S}5}\\[1mm]
\Phi^{(l-m)}_{\mathrm{S}6}\\[1mm]
\Phi^{(l-m)}_{\mathrm{S}7}\\[1mm]
\end{array}
\:\!\right], 
\label{eq:gnrl:scalar}
\end{align}
where we have used that the background reduces to the near-horizon geometry at the leading-order of $\lambda$-series. 

\par 
The structure of the above set of equations are the same as that of the Reissner--Nordstrom case. Namely, 
we have, at the leading-order, four mutually decoupled, homogeneous equations 
for $(\Phi^{0)}_{\textrm{S}1}, \: \Phi_{\textrm{S}2}^{(0)}, \: \Phi_{\textrm{S}3}^{(0)}, \: \Phi_{\textrm{S}4}^{(0)})$. 
As is also the Reissner--Nordstrom case, by using the transverse-traceless conditions~\eqref{eq:sconsfaa}, \eqref{eq:scons:HL}, and \eqref{eq:sconstrfa}, 
we find that $(\Phi_{\textrm{S}4}^{(0)}, \Phi_{\textrm{S}5}^{(0)}, \Phi_{\textrm{S}6}^{(0)}, \Phi_{\textrm{S}7}^{(0)})$ can be determined in terms of $(\Phi^{(0)}_{\textrm{S}1}, \: \Phi^{(0)}_{\textrm{S}2}, \: \Phi^{(0)}_{\textrm{S}3})$. 
Therefore we can view the three variables $(\Phi^{(0)}_{\textrm{S}1}, \: \Phi^{(0)}_{\textrm{S}2}, \: \Phi^{(0)}_{\textrm{S}3})$ as the leading-order master variables, 
governed by the homogeneous master wave equations on the near-horizon geometry, schematically expressed as 
\ben 
{\cal L}^{(0)} \Phi^{(0)}  = 0 \,. 
\label{gnrl:eqs:master:lead:scalar}
\een 
We also have, at $l(\geqslant 1)$th-order, three mutually decoupled inhomogeneous equations for $(\Phi^{(l)}_{\textrm{S}1}, \: \Phi^{(l)}_{\textrm{S}2}, \: \Phi^{(l)}_{\textrm{S}3})$, schematically expressed in the form 
\ben
{\cal L}^{(0)} \Phi^{(l)}  = S^{(l)}  \,, 
\een
where the source term $S^{(l)}$ consists only of the lower-order variables $\Phi^{(l-m)}$, 
and the remaining equations can be used to determine $\Phi_{\textrm{S}J}^{(l)}, \: J=4,5,6,7$. Therefore, as in the Reissner--Nordstrom case, 
once the leading-order master variables $\Phi_{\textrm{S}I}^{(0)},\: I=1,2,3$ have been obtained, we can successively obtain the solutions at any order of $\lambda$.

\section{Summary and Discussion}
\label{sec:6}

In this paper we have developed a new perturbation method to solve linear massive tensor (spin-2) fields $h_{\mu \nu}$ 
in a fairly generic class of $D=(2+n)$-dimensional static, extremal and near-extremal black hole spacetimes. Within our perturbation framework, we have derived, 
for the first time, a set of mutually decoupled master equations for the massive tensor field on such a black hole background. When $D \geqslant 5$, the massive tensor field can be classified into the tensor-, vector-, and scalar-type 
according to their behavior on ${\cal K}^n$, and one can treat each tensorial type separately. 
We have introduced tensor harmonics ${\mathbb T}_{ij}, {\mathbb V}_{i}$, and ${\mathbb S}$ on the Einstein space ${\cal K}^n$, separated the ``angular variables" and reduced the massive tensor field equations to the sets of equations on the $2$-dimensional spacetime ${\cal N}^2$ spanned by the Killing time and radial coordinates $y^a=(v,x)$. At this stage, the reduced equations of motion were still coupled. We then restricted our attention to the extremal and near-extremal 
black hole backgrounds, which admit the near-horizon geometry obtained by taking the scaling limit $\lambda \rightarrow 0$. 
Then, we performed the near-horizon expansion of both the background metric and the field variables with respect to $\lambda$ and derived the set of 
mutually decoupled master equations on ${\cal N}^2$: Eq.~(\ref{eq:gnrl:tensor}) for the tensor-type, Eqs.~(\ref{eq:gnrl:vector}) for the vector-type, and Eqs.~(\ref{eq:gnrl:scalar}) for the scalar-type. At each order of $\lambda$ (say, $l$-th order), the tensor-type has the single master variable $\Phi^{(l)}_{\rm T}$, the vector-type has the two master variables $(\Phi^{(l)}_{{\rm V}1}, \Phi^{(l)}_{{\rm V}2})$, which can determine the remaining one $\Phi^{(l)}_{{\rm V}3}$, and the scalar-type has the three master variables $(\Phi^{(l)}_{\textrm{S}1}, \: \Phi^{(l)}_{\textrm{S}2}, \: \Phi^{(l)}_{\textrm{S}3})$, which can determine the remaining four variables $\Phi^{(l)}_{\mathrm{S}J}, \: J=4,5,6,7$. 
Thus, we obtained the six master variables $\Phi^{(l)}_{\rm T}$, $(\Phi^{(l)}_{{\rm V}1}, \Phi^{(l)}_{{\rm V}2})$, and $(\Phi^{(l)}_{\textrm{S}1}, \: \Phi^{(l)}_{\textrm{S}2}, \: \Phi^{(l)}_{S3})$. 
Collectively denoting these master variables by $\Phi^{(l)}$, we have found that the massive tensor field equations are reduced to the decoupled set of equations, schematically expressed in the form 
\ben
   {\cal L}^{(0)}  \Phi^{(l)}  = S^{(l)} \,,  
\label{eq:gnrl:all}
\een
where ${\cal L}^{(0)}$ is the derivative operator on the near-horizon geometry ${\cal N}^2$, and where the source term $S^{(l)}$ consists only of the lower-order variables 
$\Phi^{(l-m)}$. For the leading-order $l=0$, this reduces to the homogeneous wave equations with vanishing source term. Thus, once we have solved 
the leading-order master equations, by constructing Green's function of ${\cal L}^{(0)}$, we can successively solve the above equations (\ref{eq:gnrl:all}). 
Together with the tensor harmonics ${\mathbb T}_{ij}, {\mathbb V}_{i}$, and ${\mathbb S}$ on ${\cal K}^n$, the master variables $\Phi^{(l)}=\{
\Phi^{(l)}_{\rm T}, \Phi^{(l)}_{{\rm V}1}, \Phi^{(l)}_{{\rm V}2}, \Phi^{(l)}_{\textrm{S}1}, \: \Phi^{(l)}_{\textrm{S}2}, \: \Phi^{(l)}_{\textrm{S}3} \}$ on ${\cal N}^2$ describe the all $n(n+3)/2$ dynamical 
degrees of freedom for $h_{\mu \nu}$. 
As a concrete case, we have performed the above analysis in the four-dimensional extremal and near-extremal Reissner--Nordstrom black hole background. 
In this four-dimensional background, we have only the vector- and scalar-type components as the tensor-type components become trivial. Thus, we have 
five master scalar variables, $\Phi^{(l)}=\{ \Phi^{(l)}_{{\rm V}1}, \Phi^{(l)}_{{\rm V}2}, \Phi^{(l)}_{\textrm{S}1}, \: \Phi^{(l)}_{\textrm{S}2}, \: \Phi^{(l)}_{\textrm{S}3} \}$ at each order. We have given the general solutions for the leading order master equations~(\ref{eqs:master:lead:vector}) and (\ref{eqs:master:lead:scalar}). 

\par
In this paper, we focused on formulating our perturbation method. 
For applications in astrophysical problems, clearly an important task is to generalize the present method to the extremal and near-extremal Kerr black hole background. Another interesting problem may be 
to use our perturbation method to study in detail the stability of static extremal and near-extremal black holes in asymptotically AdS spacetimes in the context of 
AdS-CFT correspondence.

\begin{acknowledgments}
We thank the Yukawa Institute for Theoretical Physics (YITP), at Kyoto University, where this work was initiated during 
the YITP-T-18-05 workshop on {\em Dynamics in Strong Gravity Universe}.  
This work was supported by the MEXT-Supported Program for the Strategic Research Foundation at Private Universities, 2014--2017 (S1411024)~(T.~I.) and JSPS KAKENHI Grant Numbers 15K05092~(A.~I.) and JP19K14715~(T.~I.). 
%
%
V.C.\ acknowledges financial support provided under the European Union's H2020 ERC 
Consolidator Grant ``Matter and strong-field gravity: New frontiers in Einstein's 
theory'' grant agreement no. MaGRaTh--646597.
This project has received funding from the European Union's Horizon 2020 research and innovation programme under the Marie Sklodowska-Curie grant agreement No 690904.
We acknowledge financial support provided by FCT/Portugal through grant PTDC/MAT-APL/30043/2017.
The authors would like to acknowledge networking support by the GWverse COST Action 
CA16104, ``Black holes, gravitational waves and fundamental physics.''

\end{acknowledgments}

\section*{Appendix: Concrete expressions of $\Delta^{(m)}_{(abc;def)}$} 

In this appendix, we provide, for some typical cases, the concrete expressions of $\Delta^{(m)}_{(abc;def)}$ introduced by Eq.~(\ref{def:Delta:abcdef}),  
\begin{align}
&r^a r'^br''^c g^{d} g'^e g''^f=\sum_{m=0}^\infty \lambda^m \cdot \Delta^{(m)}_{(abc;def)} \,. 
\end{align}
Hereafter we denote $k:=a+b+c+d+e+f$.  

\medskip 

\noindent 
$\bullet$ $r^a$ ($a\neq0$, $b=\cdots =f=0$)
\begin{align}
\Delta^{(m_0)}_{(a00;000)}
&=\sum_{m_1=0}^{m_0} \cdots \sum_{m_{k-1}=0}^{m_{k-2}}r^{(m_0-m_1)}\cdots r^{(m_{k-2}-m_{k-1})}\:\!r^{(m_{k-1})}
\nonumber \\
&=\sum_{m_1=0}^{m_0} \cdots \sum_{m_{k-1}=0}^{m_{k-2}} \left[\:\!
\prod_{p=1}^{a-1} r^{(m_{p-1}-m_p)}
\:\!\right]r^{(m_{k-1})}.
\end{align}

\medskip

\noindent
$\bullet$ $r^a r'^b$ ($b\neq0$, $c=\cdots=f=0$)
\begin{align}
\Delta^{(m_0)}_{(ab0;000)}
&=\sum_{m_1=0}^{m_0} \cdots \sum_{m_{k-1}=0}^{m_{k-2}}
\left[\:\!
r^{(m_0-m_1)}\cdots r^{(m_{a-1}-m_a)}\:\!\right]\left[\:\!
r'^{(m_a-m_{a+1})}\cdots r'^{(m_{a+b-2}-m_{a+b-1})}\:\!\right]r'^{(m_{k-1})}
\nonumber \\
&=\sum_{m_1=0}^{m_0} \cdots \sum_{m_{k-1}=0}^{m_{k-2}}
\left[\:\!
\prod_{p=1}^a r^{(m_{p-1}-m_p)}
\:\!\right]\left[\:\!
\prod_{q=1}^{b-1} r'^{(m_{a+q-1}-m_{a+q})}
\:\!\right] r'^{(m_{k-1})}.
\end{align}

\medskip

\noindent
$\bullet$ $r^a r'^br''^c$ ($c\neq0$, $d=\cdots=f=0$)
\begin{align}
\Delta^{(m_0)}_{(abc;000)}
=\sum_{m_1=0}^{m_0} \cdots \sum_{m_{k-1}=0}^{m_{k-2}}\left[\:\!
\prod_{p=1}^a r^{(m_{p-1}-m_{p})}
\:\!\right]\left[\:\!
\prod_{q=1}^b r'^{(m_{a+q-1}-m_{a+q})}
\:\!\right]\left[\:\!
\prod_{s=1}^{c-1} r''^{(m_{a+b+s-1}-m_{a+b+s})}
\:\!\right] r''^{(m_{k-1})}.
\end{align}

\medskip

\noindent
$\bullet$ 
$r^ar'^b r''^c g^d$ ($d\neq0$, $e=f=0$)
\begin{align}
\Delta^{(m_0)}_{(abc;d00)}
=&\sum_{m_1=0}^{m_0} \cdots \sum_{m_{k-1}=0}^{m_{k-2}}\left[\:\!
\prod_{p=1}^a r^{(m_{p-1}-m_{p})}
\:\!\right]\left[\:\!
\prod_{q=1}^b r'^{(m_{a+q-1}-m_{a+q})}
\:\!\right]\left[\:\!
\prod_{s=1}^{c} r''^{(m_{a+b+s-1}-m_{a+b+s})}
\:\!\right]
\cr
&\times
\left[\:\!
\prod_{t=1}^{d-1} g^{(m_{a+b+c+t-1}-m_{a+b+c+t})}
\:\!\right]g^{(m_{k-1})}.
\end{align}

\medskip

\noindent
$\bullet$ $r^ar'^b r''^c g^d g'^e$ ($e\neq0$, $f=0$)
\begin{align}
\Delta^{(m_0)}_{(abc;de0)}
=&\sum_{m_1=0}^{m_0} \cdots \sum_{m_{k-1}=0}^{m_{k-2}}\left[\:\!
\prod_{p=1}^a r^{(m_{p-1}-m_{p})}
\:\!\right]\left[\:\!
\prod_{q=1}^b r'^{(m_{a+q-1}-m_{a+q})}
\:\!\right]\left[\:\!
\prod_{s=1}^{c} r''^{(m_{a+b+s-1}-m_{a+b+s})}
\:\!\right]
\cr
&\times
\left[\:\!
\prod_{t=1}^{d} g^{(m_{a+b+c+t-1}-m_{a+b+c+t})}
\:\!\right]\left[\:\!
\prod_{u=1}^{e-1} g'^{(m_{a+b+c+d+u-1}-m_{a+b+c+d+u})}
\:\!\right]g'^{(m_{k-1})}.
\end{align}

\medskip

\noindent
$\bullet$ $r^ar'^b r''^c g^d g'^eg''^f$ ($f\neq0$)
\begin{align}
\Delta^{(m_0)}_{(abc;def)}
=&\sum_{m_1=0}^{m_0} \cdots \sum_{m_{k-1}=0}^{m_{k-2}}\left[\:\!
\prod_{p=1}^a r^{(m_{p-1}-m_{p})}
\:\!\right]\left[\:\!
\prod_{q=1}^b r'^{(m_{a+q-1}-m_{a+q})}
\:\!\right]\left[\:\!
\prod_{s=1}^{c} r''^{(m_{a+b+s-1}-m_{a+b+s})}
\:\!\right]
\cr
&\times
\left[\:\!
\prod_{t=1}^{d} g^{(m_{a+b+c+t-1}-m_{a+b+c+t})}
\:\!\right]\left[\:\!
\prod_{u=1}^{e} g'^{(m_{a+b+c+d+u-1}-m_{a+b+c+d+u})}
\:\!\right]
\cr
&\times\left[\:\!
\prod_{v=1}^{f-1} g''^{(m_{a+b+c+d+e+v-1}-m_{a+b+c+d+e+v})}
\:\!\right]
g''^{(m_{k-1})}.
\end{align}


\end{document}